\documentstyle[aps,12pt,epsfig]{revtex}

\newcommand{\beq}{\begin{equation}}

\newcommand{\eeq}[1]{\label{#1} \end{equation}}

\newcommand{\beqar}{\begin{eqnarray}}

\newcommand{\eeqar}[1]{\label{#1} \end{eqnarray}}

\newcommand{\bmath}{\begin{displaymath}}

\newcommand{\emath}{\end{displaymath}}

\newcommand{\bitem}{\begin{itemize}}

\newcommand{\eitem}{\end{itemize}}

\begin{document}

\normalsize

%\begin{center}

%%\begin{flushright}

%%{\large \bf MG2004-PHY-11}

%%\end{flushright}

\title{\Large \bf Baryon Junction Loops in HIJING/B\=Bv2.0
and the Baryon/Meson Anomaly at RHIC } 
\author{~V. Topor Pop$^{1}$,  M. Gyulassy$^2$, 
J. Barrette$^{1}$, C. Gale$^{1},$ \\
~X. N. Wang$^3$ and N. Xu$^3$}

\address{
$^1$McGill, University, Montreal, Canada, H3A 2T8\\
$^2$Physics Department,
Columbia University, New York, N.Y. 10027\\
$^3$Nuclear Science Division, LBNL, Berkeley, CA 94720\\
}

\date{July 26, 2004}

\maketitle

%\vskip 0.3cm

\begin{abstract}
A new version, v2.0, of the HIJING/B\=B Monte Carlo
nuclear collision event generator 
is introduced in order to explore further
the possible role of baryon junctions 
 loops in the baryon/meson anomaly (2 $< p_{T} <$ 5 GeV/c) observed
in 200A GeV Au+Au reactions at RHIC.
We show that junction loops with  an enhanced
intrinsic $k_T\approx 1$ GeV/c transverse momentum kick
may provide a partial explanation of the anomaly
as well as other important baryon stopping observables.
\end{abstract}

\newpage
\section{Introduction}

The phase transition from partonic degrees of freedom (quarks and
gluons) in ultra-relativistic nuclear collisions 
to hadronic degree of freedom is a central focus of
recent experiments at RHIC \cite{qm02,qm04,gyu_qm04,bnl04,gyu_bnl04,muller04}.
One of the interesting and unexpected 
discoveries\cite{Adcox:2001mf,Adler:2003kg} at RHIC is the 
"baryon anomaly" \cite{Vitev:2001zn}, observed as 
a large enhancement of the baryon to meson ratio
and as a large difference between the nuclear modification
factor $R_{AA}(p_T)$ between total charged and $\pi^0$ 
at moderate  $2<p_T<5$ GeV/c. 
There are two main effects that contribute to this anomaly.
One is the predicted jet quenching\cite{Vitev:2001zn,Gyulassy:2003mc} 
that strongly
suppresses the pion yield above $p_T>2$ GeV/c. This effect causes an 
apparent enhancement
since the pion denominator is reduced. The second effect is 
a genuine  enhancement of baryon transverse momentum spectra.
Several effects can contribute to such enhancements of baryon
yields at moderate $p_T$. Radial hydrodynamical flow
has been observed at all
energies including RHIC\cite{Xu:2001zj}. However at high $p_T$ 
local equilibrium
must fail. At intermediate $p_T$ a nonequilibrium remnant of hydrodynamic
flow may arise from multi quark recombination\cite{muller04,lin02}.

In this work we continue to explore  the possibility that a 
more novel and unconventional
source of baryon production may be at least partially responsible for
the baryon anomaly. We study whether  baryon junction loops, as proposed
in \cite{svance99} to explain (anti)hyperon 
production at lower (SPS) energies, could help explain the RHIC data.
The idea that nonperturbative three color flux junctions
could play an important role in baryon and anti-baryon production
at high energies was proposed long ago by 
Rossi and Veneziano\cite{Rossi:1977cy,rossi80} 
on the basis of dual regge theory. This idea was 
extended and applied by Kharzeev \cite{kharzeev96}
to nuclear collisions. The first full A+A event Monte Carlo  implementation 
of this mechanism was HIJING/B generator by Vance et al.\cite{svance98}. 
The addition of baryon junction loop mechanism led to 
HIJING/B\=B1.10 version of this model\cite{svance99}.
Unlike conventional diquark fragmentation models,
a baryon junction allows the diquark to split with the
three independent flux lines tied together at a junction.
 For an alternate possible formulation of baryon junction dynamics 
see \cite{Kopeliovich:1988qm,capella02,kopelovic99,bopp04}.
%%% add reference Capela, and....

In this paper we  introduce a new version (v2.0)
of HIJING/B\=B that differs from HIJING/B\=B1.10 \cite{svance99} 
in its implementation of hypothesized junction anti-junction loops
that may be responsible for novel baryon pair production in 
nucleus nucleus (A+B) collisions.

A large data base on meson and baryon spectra
\cite{Adcox:2001mf,Adler:2003kg,Adler:2003cb,Adcox:2003nr,Adams:2003xp,Adams:2003ve,Adler:2001aq,Ouerdane:2004yw,Bearden:2003hx,Lee:2003iq,Bearden:2003fw,Christiansen:2002gn,Bearden:2002ry,Bearden:2001kt,Back:2003ff,Wosiek:2002ur,Back:2002ks,Back:2001qr}
are now available from RHIC experiments.
The HIJING event generator\cite{hij92_99} was developed to 
extrapolate hadron-hadron multiparticle soft plus hard phenomenology
as encoded in the LUND JETSET/PYTHIA model\cite{Sjostrand:1993yb}
to nuclear collisions. 
One important feature of HIJING is that
it can account for the pion quenching component of the baryon anomaly.
However, the LUND JETSET di-quark string fragmentation  
mechanism used in HIJING v1.37 \cite{hij92_99} 
completely fails to describe the baryon spectra 
observed in A+A collisions at all energies 
\cite{ToporPop:1995cg,Gyulassy:1997mz,top03_prc68}. 
HIJING/B\=B1.10 was developed to address this failure
at SPS energies. It was however also found to be also inadequate, as
we review below, with respect to baryon observables.
For a recent discussion comparing HIJING1.37 and  HIJING/B\=B1.10 
predictions for global observables at RHIC see \cite{top03_prc68}.

Clarifying the physical origin of the (anti) baryon dynamics
at RHIC is important given the variety of 
 hadronization mechanisms proposed in 
hydrodynamic models \cite{heinz03},  
multi-quark coalescence\cite{lin02}, thermal models \cite{therm03},
and the novel baryon junction dynamics 
\cite{Vitev:2001zn,Sjostrand:2004pf}.
The valence baryon number migration
over the large rapidity window $-5<y<5$ at RHIC
provides another stringent  tests of the baryon dynamics.
Recent RHIC data show at midrapidity a sizeable
finite net-proton ($p-\bar{p}$) number in the final state 
\cite{Adcox:2001mf,Adler:2003kg,Adler:2003cb,Adcox:2003nr,Adams:2003xp,Adams:2003ve,Adler:2001aq,Ouerdane:2004yw,Bearden:2003hx,Lee:2003iq,Bearden:2003fw,Christiansen:2002gn,Bearden:2002ry,Bearden:2001kt,Back:2003ff,Wosiek:2002ur,Back:2002ks,Back:2001qr}.
Moreover, the antibaryon-to-baryon ratios are not equal to one,
providing further evidence for  non-transparency of high energy nuclei.
This significant baryon number transport over more than 5 units in 
rapidity has inspired other approaches as well
\cite{stop_pap_01,stop_pap_02,stop_pap_03,stop_pap_04,capella03}.
The net baryon density at midrapidity
also has an impact on the hadro-chemical equilibration affecting hadronic
yields \cite{munzinger99}.
% and the build up of collective flow 
%\cite{greiner74}.

Another possible source of novel baryon/hyperon physics
are strong color electric fields (SCF). This is modeled as
an increase of the effective  string tension that controls the
$q\bar{q}$ and $qq\bar{qq}$ pair creation rates
\cite{ropes_tbiro,ropes_sorge}.
Molecular dynamics model \cite{qmd_1,qmd_2,urqmd_1,urqmd_2}
have been used to study the effects of color ropes
as an  effective description of the
non-perturbative, soft gluonic part of QCD \cite{soff_jpg04,nu03_scf,nu04_scf}.
%An observables which has been shown to be highly sensitive to SCF 
%is the net-baryon number transport.
SCF increases the partons {\em ``intrinsic transverse momentum''} 
($k_T$) and decreases the protons and antiprotons yields. 
The empirical value of the Regge slope for baryon   
is  $ \alpha\,' \simeq 1\,\,GeV^{-2}$ that 
yields a string tension  $\kappa$ 
( related to the Regge slope, $\kappa=1/2 \pi \alpha'$),
of approximately 1 GeV/fm.
It has been suggested that the magnitude of a typical
field strength at RHIC energies might reach 5-12 GeV/fm 
\cite{csernai01}.
However, in this work we will not consider further the consequences of
SCF effects but concentrate only on the junction loops.

%Present experimental evidence \cite{pdbook_02} indicates
%that baryons can be produced only in particle-antiparticle pairs. 
%Therefore, in addition we modified the algorithm logic, to
%allow multiple J\=J loops introducing now a recurence formula 
%for the calculations of the probabilities that 
%a given diquark or antidiquark gets an ``{\em enhanced $p_{T}$
%kick''} from the enderlying junction mechanism.

In the following sections we compare numerical results
of HIJING/B\=B v2.0
to transverse momentum spectra,
rapidity densities (dN/dy) of protons (p), anti-protons (\=p),
and net-protons (p-\=p) for central Au+Au collisions
at $\sqrt{s_{NN}}$=200 GeV as well as their centrality dependence. 
The characteristic stopping observables: 
average rapidity loss, energy loss of net-baryons
per participant nucleon 
%(see section III for definitions) 
and transverse energy per net-baryons are presented.
The ``anomalous'' baryon-mesons composition 
at moderate $p_{T}$ observed  
in particle ratios $\bar{p}/p$, $p/\pi^{+}$, $\bar{p}/\pi^{-}$
and in species dependent nuclear modification factors, 
is also discussed.

In Section II we briefly recall the ingredients of the 
HIJING \cite{hij92_99} , HIJING/B\=B (v1.10) \cite{svance98,svance99} 
approaches and point out the 
extensions incorporated in the new version HIJING/B\=B v2.0.
In Section III numerical results are discussed in detail 
in comparison with available data. 
Summary and Conclusions are presented in Section IV.

\section{ HIJING/B\=B v2.0.}

HIJING is a model that provides a theoretical framework
to extrapolate elementary proton-proton multiparticle phenomena 
to complex nuclear collisions
as well as to explore possible new physics such as energy loss and gluon 
shadowing\cite{stop_pap_04}. 
Three versions of this model will be compared in this paper: HIJING v1.37,
HIJING/B\=B v1.10 and HIJING/B\=B v2.0.
Detailed descriptions of the  HIJING v1.37 and  HIJING/B\=B v1.10
models can be found in 
Refs. \cite{svance99,svance98,hij92_99,top03_prc68}.

In HIJING1.37  \cite{hij92_99} 
the soft beam jet fragmentation is modeled by diquark-quark strings 
as in \cite{Sjostrand:1993yb} with gluon
kinks induced by soft gluon radiations. 
The mini-jets physics 
is computed via an eikonal multiple collision framework
using pQCD PYTHIA5.3 to compute the initial and final state radiation
and hard scattering rates. 
The cross section for hard parton scatterings is enhanced by
a factor K=2 in order to simulate high 
order corrections. HIJING extends PYTHIA to 
include a number of new nuclear effects.
Besides the Glauber nuclear eikonal extension,
shadowing of nuclear parton distributions
is modeled. In addition  dynamical energy loss
of the (mini)jets is taken into account through an effective dE/dx.

In HIJING/B\=B v1.10 \cite{svance99}
the baryon junction mechanism was introduced as an extension
of HIJING/B\cite{svance98} in order to try to account
for the observed longitudinal distributions of 
baryons(B) and anti-baryons(\=B) in proton nucleus (p+A) and 
nucleus-nucleus (A+A) collisions at the SPS energies.
However, as implemented in  HIJING/B\=Bv1.10 the junction loops
still  fail to account 
for the observed transverse slopes of anti-baryons at moderate $p_T$ 
as shown in \cite{top03_prc68}. This motivated us to try
to reformulate the junction loop implementation in the present
HIJING/B\=B (v2.0).

The  J\=J loop algorithm of v1.10 has been replaced
by a simple enhancement of the intrinsic (anti)diquark
$p_T$ kicks in any string that has been selected
to contain one or more loops. Multiple hard and soft interactions 
proceed as in HIJING1.37.
Before fragmentation however via JETSET we compute the probability that
a junction loop occurs in the string.
A picture of a juction loop is as follows: a
color flux line splits at some intermediate point
into two flux line at one junction and then the flux line fuse back
into one at a second anti-junction somewhere further 
along the original flux line. 
The distance in rapidity between these points is chosen
via a Regge distribution as described bellow.
For single inclusive baryon observables this distribution does not need
to be specified.

The probability of such a loop is assumed to
increase with the number of binary interactions, $n_{hits}$ that  the
incident baryon suffers in passing through the oncoming nucleus.  This number
depends on the relative and absolute impact parameters 
and is computed in HIJING
using the eikonal path through a diffuse nuclear density.

We assume as in \cite{svance99} that out of the non single diffractive
NN interaction cross section, $\sigma_{in}-\sigma_{nsd}$,
a fraction $f_{J\bar{J}}=\sigma_{J \bar{J}}/(\sigma_{in}-\sigma_{nsd})$
of the events excite a junction loop.
The probability after $n_{hits}$ that the incident baryon
has a $J\bar{J}$ loop is:
%\bmath
\begin{equation}
P_{J\bar{J}}=1-(1-f_{J\bar{J}})^{n_{hits}}
\end{equation}
%\emath
We take $\sigma_{J \bar{J}}$=17 mb, $\sigma_{sdf}\,\, \approx$ 10 mb
and the total inelastic nucleon nucleon cross 
section $\sigma_{in}\,\, \approx$ 42 mb at RHIC energies.
These cross sections imply that a junction loop occurs in pp collisions
at RHIC energy with a rather high probability $17/32\approx 0.5$
and rapidly approaches 1 in $AA$. In $p+S$ where $n_{hit} \approx 2$
there is an 80\% probability that a junction loop occurs in this scheme.
Thus the effects of loops is taken here to have a very rapid onset and
essential all participant baryons are excited with $J\bar{J}$ loops
in $AA$ at RHIC.
We investigate the sensitivity of the results to the value of parameter   
$\sigma_{J \bar{J}}$ and found no significant variation 
on pseudo-rapidity distributions of charged particles, 
for 15 mb $< \sigma_{J \bar{J}} <$ 25 mb for Au+Au. 
Light ion reactions like $p+S$ and $p+Ar$ would have more sensitivity
to $\sigma_{J \bar{J}}$.

The production of a baryon and antibaryon from a $J\bar{J}$ loop is
simulated via an enhancement of the diquark $p_T$ kick parameter
$\sigma_{qq}=PARJ(21)$ of JETSET7.3. The default value is 
$\sigma_{qq}$ = 0.36 GeV/c in ordinary string fragmentation.
However in events where the string has a junctions loop
we can expect a significantly higher $p_T$ kick \cite{svance99}.
We therefore propose a very simple algorithm whereby
the $J\bar{J}$ is modeled by enhancing $\sigma_{qq}$
by a factor $F_{p_T}$ which we fit to best reproduce the observed
$p_T$ spectrum of the baryons. This implementation of
the $J\bar{J}$ model marks a radical departure 
from that implemented in HIJING/B\=Bv1.10.

While the above model allows the baryon antibaryon pairs to acquire much 
high transverse momentum in accord with observation,
the absolute production rate also depends on the diquark/quark 
suppression factor $PARJ(1)$. 
The JETSET default for ordinary (fundamental flux) strings 
has $PARJ(1)=0.1$. The reduced  number
of protons and anti protons observed at RHIC relative to HIJING1.37, 
will be shown below to be accounted for, if PARJ(1)  is reduced
 to 0.07  in $J\bar{J}$ loops.

In summary two paramters, $PARJ(1), PARJ(21)$, are used in version 2.0
to {\em simulate} the dynamical consequences of the hypothesized $J\bar{J}$ 
loop production in $A+B$ reactions.  
The factor $F_{p_T}$ modifying the default 0.36 GeV value of $PARJ(21)$
may depend on beam energy, 
atomic mass number (A), centrality (impact parameter).
However, we will show that a surprising good description of a variety
of observables is obtained with a constant value, $F_{p_T}$=3. 
The sensitivity of the 
theoretical predictions to this parameter is discussed in section III.

Finally, we remark that correlations studies 
in p+p and p(d)+Au collisions 
at RHIC energies could eventually help us to obtain
more precise values of J\=J loop parameters 
(mainly: $\sigma_{J\bar{J}}$, Regge intercept $\alpha_{J}(0)$ and $F_{p_T}$).
The contribution to the double differential inclusive cross
section for the inclusive production of a $B$ and
$\bar{B}$ in NN collisions due to $J\bar{J}$ exchange
is \cite{svance99,kharzeev96}:

%\bmath

\begin{equation}
E_BE_{\bar{B}} \frac{d^6 \sigma } {d^3 p_B d^3 p_{\bar{B}}}
\rightarrow C_{B\bar{B}}e^{(\alpha_{J}(0)-1)|y_{B}-y_{\bar{B}}|}
\end{equation}
%\emath
where $C_{B\bar{B}}$ -is an unknown function of the transverse
momentum and  $M_0^J+ P +B$ (junction+Pomeron+baryon) couplings
\cite{svance99,svance98}.
The predicted rapidity correlation length 
($1-\alpha_{J}(0))^{-1}$
depends upon the value of the intercept $\alpha_{J}(0)$.
To test for  $M_0^J$ component $\alpha_{J}(0) \simeq 0.5$ 
requires the measurement of rapidity correlations on 
a scale $|y_B-y_{\bar{B}}| \sim 2$.
In contrast, infinite range rapidity correlations are suggested if 
$\alpha_{J}(0) \simeq 1.0$ \cite{kopelovic99}.
It is thus important to look for rapidity correlations  
at RHIC energies where very high statistics data are now 
available, in  $p + p \rightarrow B + \bar{B} + X$,
or $p(d) + A \rightarrow B + \bar{B} + X$.

\section{Numerical Results}

\subsection{Transverse Momentum Spectrum}

The nuclear modification factor ($R_{AA}$) is defined as the ratio of the
hadron yield in central Au+Au collisions to that in 
p+p reactions scaled by the number of binary collisions ($N_{coll}$):

\begin{equation}
R_{AA}(p_{\perp})=\frac{d^2N_{AA}/dydp_{\perp}}
{<N_{coll}>d^2N_{pp}/dydp_{\perp}}
\end{equation}
where, $<N_{coll}>$ is the average number of binary collisions of the
event sample calculated from the nuclear overlap integral
($T_{AA}$) and the inelastic nucleon-nucleon cross section;\\ 
$<N_{coll}>= \sigma^{inel}_{nn}<T_{AA}>$.

In Fig.~\ref{fig:fig01_st} the measured 
\cite{Adler:2003au} nuclear modification 
factor ($R_{AA}$) for charged hadrons 
in central (0-10\%) Au+Au collisons at 200 GeV 
are compared to the predictions of HIJING v1.37 and 
HIJING/B\=B v2.0 models. 
The data shows the strong jet quenching effect that suppresses the 
hadrons yield by a factor of $ \approx 5$ for the highest $p_T$ bins
resulting in an observed maximum in $R_{AA}$ 
at $p_T \approx $ 2 GeV/c.  
Note that both HIJING v1.37 and HIJING/B\=B v1.10 
fail to reproduce
the ``baryon bump'' at moderate $p_T$ seen in the $R_{AA}$
factor and also fail to account 
for the large transverse slopes of baryons and anti-baryons
(see ref. \cite{top03_prc68} ).     
The Lund string fragmentation mechanism of hadronization in HIJING
v1.37 leads to a rather slow increase of 
the nuclear modification factor $R_{AA}$ to unity at high
$p_T$, not observed in the data (see part a, results 
without quenching and shadowing effects (nqs)).
The addition of jet quenching and shadowing effects (yqs) in HIJING v1.37
still fails to describe the data. 
%in particular the 
%observed distinct localized bump at $p_T \approx $ 2 GeV/c.

In contrast, HIJING/B\=B v2.0 with   
shadowing and jet quenching (yqs) effects with default 
energy loss parameter dE/dx=1 GeV/fm (for quark jet)
describes well the data over the full $p_T$ range.
Some of the observed discrepancies 
could be attributed to strange and multistrange
hyperons which are underestimated in our calculations
because we do not consider SCF effects here.

Figure ~\ref{fig:fig02_st} presents a comparison of 
the experimental transverse mass distributions \cite{Adler:2003cb} 
of positive (left) and negative (right) particles with the 
predictions of HIJING/B\=B v2.0 (upper panel) 
and HIJING v1.37 (lower panel).
The data shows a mass dependence in the shape of the spectra.
The protons (p) and anti-protons (\=p) spectra have a shoulder-arm
shape at low $p_T$ characteristic of a radial flow.
%We mention that the data \cite{Adler:2003cb}
%for p and \=p are feeddown corrected.
The pion spectra are well described by both models.
Introducing corrected J\=J loops algorithm in 
HIJING/B\=B v2.0 result in a significant improvement in the description of 
the protons and anti-protons in the scenario 
with shadowing and jet quenching. 
However, only a qualitative description is obtained for low $m_T-m_0$
spectra due to the presence of radial flow,
not included in the model.
A similar conclusion can be drawn from the predictions 
of the models for mean transverse momenta.

\subsection{ Stopping Observables}

Rapidity distributions of participants (net) baryons are very
sensitive to the dynamical and statistical properties of 
nucleus-nucleus collisions. 
The RHIC net proton distribution is both qualitatively and
quantitatively different from those at lower 
AGS and SPS energies \cite{Bearden:2003hx}.
Recent results for net-proton in central (0-5\%) Au+Au interactions at 
total nucleon nucleon centre of mass energy  
$\sqrt{s_{NN}}$=200 GeV show an unexpectedly large rapidity
density at midrapidity \cite{Adcox:2003nr,Bearden:2003hx}.

Figure ~\ref{fig:fig03_st} presents a comparison of
the rapidity distributions of protons (Fig.~\ref{fig:fig03_st}a), 
anti-protons (Fig.~\ref{fig:fig03_st}b) and 
net-protons (p-\=p) (Fig.~\ref{fig:fig03_st}c) 
and their ratio (p/\=p) (Fig.~\ref{fig:fig03_st}d)
obtained for Au+Au interactions  
at total c.m energy $\sqrt{s}$=200A GeV with the model predictions 
of HIJING/B\=B v2.0 and RQMD v2.4 \cite{ropes_sorge}. 
Corrections for feed-down contributions 
have been applied to the data. 
We discuss here a comparison with RQMD v2.4 results 
in order to investigate if
hadronic rescattering only and SCF effects as implemented in RQMD,
could explain the new data.
The new version of HIJING/B\=B reproduces very well the 
experimental yield at mid-rapidity for both 
p and \=p as well as their ratio and the net protons yield (p-\=p). 
This agreement is improved if shadowing and jet quenching are included.
In contrast RQMD v2.4 does not reproduce the shape of the proton 
rapidity distribution near midrapidity
and strongly underpredicts the anti-proton yield.

In addition, centrality dependence of proton and anti-proton yields 
at mid-rapidity have been also analysed
and the results are shown in Fig.~\ref{fig:fig04_st}.
HIJING/B\=B v1.0 overpredicts the data \cite{Adcox:2003nr}
except for very peripheral collisions.
In contrast, HIJING/B\=B v2.0 reproduces very well the 
experimental yield at all centralities.

One of the main feature of the data is the observed increase of
net-proton up to three units of rapidity away from midrapidity (Fig. 3c).
This central valley could be used as an indicator for 
partonic processes \cite{bass_st_03}, \cite{wolschin_03}.
Microscopically, the baryon number transport over 4-5 units of rapidity
to the equilibrated midrapidity region is not only due to
hard processes acting on single valence (di)quark that are 
described by perturbative QCD, since this yields 
insufficient stopping \cite{ToporPop:1995cg},{\cite{Gyulassy:1997mz}.
Instead, additional processes such as the nonperturbative 
junction mechanism as implemented in  HIJING/B\=B v2.0
are able to reproduce the observed distribution.
Such mechanism may lead to substantial stopping even at LHC energies.

The net-baryon (B-\=B) distribution retains information about the
energy loss and allows the degree of nuclear stopping to be 
determined. Experimentaly, to obtain the net-baryons, the number of
net-neutrons and net-hyperons have to be estimated. In addition, 
the data need to be extrapolated to full rapidity space 
introducing other systematic errors. In contrast,
in models we can calculate directly specific
stopping observable as the average rapidity loss 
and the energy loss per participant nucleon as defined in
ref. \cite{Bearden:2003hx}.
The average rapidity loss is defined as $<\delta y>=y_{p}-<y>$
where $y_{p}$ is the rapidity of the incoming projectile 
and $<y>$ is the mean net-baryon rapidity after the collision:

%\bmath

\begin{equation}
<y>= \frac{2}{N_{part}} \int_0^{y_{p}} y \cdot \frac{dN_{(B-\bar{B})}}{dy}\,\cdot dy
\end{equation}
%\emath
where $N_{part}$ is the number of participating nucleons in the
collision.

The total energy $E_{tot}$ per net-baryon after the collisions can be
derived using the relation:

%\bmath

\begin{equation}
E_{tot}=\frac{1}{N_{part}} \int_{-y_{p}}^{y_{p}}<m_T> \cdot cosh y
\cdot \frac{dN}{dy} \cdot dy 
\end{equation}
%\emath 
where $<m_T>=<\sqrt{p_{T}^2+m_{0}^2}>$ is the average  transverse mass.
The energy loss per participant pair could be estimated
as $\Delta E=(100.0 - E_{tot}$) GeV. 
The predictions of the models for these quantities 
are presented in Fig.~\ref{fig:fig05_st}.

Fig.~\ref{fig:fig05_st}a shows the net-baryons 
model predictions in comparison with BRAHMS data \cite{Bearden:2003hx} 
obtained from the measured net-proton
($N_{(B-\bar{B})}$ $\approx$ $2 N_{(p-\bar{p})}$).
At RHIC energies a broad minimum has developed at mid-rapidity
for net-baryon spanning
few units of rapidity indicating that collisions are 
quite transparent. 
The average net-baryons rapidity loss deduced by BRAHMS 
$<\delta y>= 2.0 \pm 0.2$ \cite{Bearden:2003hx}
is well reproduced by  HIJING/B\=B v2.0. 
In contrast, the experimental value for 
the total energy per net-baryon after the collisions,
$E_{tot}^{exp}=28 \pm 6$ GeV \cite{Bearden:2003hx}, 
are far from our theoretical predictions of both HIJING models, 
$E_{tot}\approx$ 40 GeV.
%This may be likely due to a sudden enhancement 
%in the number of degree of freedom
%as encountered in the deconfinement of participant partons
%\cite{wolschin_03}.

%%%%%%%%%%%%%%RQMD comment%%%%%%%%%%%%%

The RQMD model \cite{ropes_sorge} with rescattering and SCF effects 
strongly overpredicts the mid-rapidity net-baryon distributions and the
average rapidity loss. Partly as a consequence
$E_{tot}$ per baryon after the collisions is underpredicted.
The extrapolation to high rapidity used in \cite{Bearden:2003hx}
to obtain their values of $<\delta y>$ and $E_{tot}$ may explain part of
the observed discrepencies. 
%It seems that the experimental results 
%for both method of extrapolation applied to data
%\cite{Bearden:2003hx} contradict the model predictions especially
%for the energy loss per participant nucleon ($\Delta E=100.0-E_{tot}$).
A precise measurement of transverse energy per baryon 
(as shown in Fig. 5c) could help also in the study  
concerning the origin of rescattering 
(which could influence the dynamics of the reaction 
at hadronic or partonic stage).
However, futher analysis and  baryon 
rapidity distribution measurements at large
rapidity are needed in order to draw a final conclusion
and to use these observables as a signature for partonic processes
and for quark-guon plasma (QGP) formation.
%{\bf say something about fig 5c}

\subsection{Particle Ratios versus $p_T$}

In Figure~\ref{fig:fig06_st_cp} the $\bar{p}/\pi^-$ ratio are shown as
function of transverse momentum ($p_T$) for central (0-10 \%) (upper
part) and peripheral (60-90 \%) collisions (lower part).
The measured ratios increase at low $p_T$ and saturate at 
different values  which strongly increase from peripheral 
to central collisions.
For central collisions at moderate $p_T$ ($2<p_T<5$ GeV/c ),
the yields of anti-protons (\=p) are comparable to that of pions.
In hard scattering processes described by pQCD 
and implemented in HIJING, the ratio $\bar{p}/\pi^-$
are determined by the fragmentation of energetic partons,
independent of the initial colliding system. Therefore
within the errors those ratios are well described for 
peripheral collisions by both version of the models. 
%and the results are also in agreement
%with those obtained in $p+p$ and $e^++e^-$ collisions.
As expected in peripheral collisions there is little sensitivity
to the new ingredients implemented in HIJING/B\=B v2.0.
In contrast the clear increase in 
the $\bar{p}/\pi^-$ ratio at moderate $p_T$ 
from peripheral to central collisions, is seen to be sensitive
to the new physics 
as implemented in HIJING/B\=B v2.0. HIJING v1.37 strongly underpredicts 
the observed ratios at moderate $p_{T}$, an effect that 
is corrected in HIJING/B\=B v2.0 that includes  
an improved simulation of moderate pt junction loop with a 
factor $F_{p_T}$=3.

A similar conclusion can be drawn from the results 
obtained for $\bar{p}/p$ ratio 
for central (0-10 \%) Au+Au collisions at 200 GeV 
presented in Fig.~\ref{fig:fig07_st}
and for the $p/\pi^+$ ratio (not presented here).
These results show the significant contributions of protons
and anti-protons yields to the total particle
composition in this moderate $p_T$ region ($2<p_T< 5\,\,$ GeV/c). 
An alternative interpretation of the observed increase with centrality
is provided by parton recombination and 
fragmentation model \cite{lin02} while the hydrodynamics
\cite{heinz03} and thermal model calculations predict 
that $\bar{p}/\pi^-$ ratio exceeds unity for central collisions.

\subsection{Nuclear modification factor versus $p_T$}

In order to better quantify the particle composition at moderate $p_T$ 
we investigate the binary collision scaling of $p_T$ spectra for
charged pions and protons (anti-protons). 
Figure~\ref{fig:fig08_st} shows the predicted nuclear modification factors 
$R_{AA}$  and $R_{cp}$ for sum of protons and anti-protons (p+\=p)/2, 
and charged pions in comparison with 
available PHENIX data \cite{Adcox:2003nr}.

%Similarly to $R_{AA}$ one can consider 
%an impact parameter dependent nuclear modification factor 

$R_{cp}$ is defined as the scaled yield  ratio at different 
centrality such as the ratio of central to peripheral yield:

\begin{equation}
R_{cp}(p_{\perp})=
\frac{Yield(Central)/<N_{coll}(Central)>}{Yield(Periph.)/<N_{coll}(Periph.)>}
\end{equation}

where, $Yield$ = ($1/N_{events}$)($1/2\pi\,p_{\perp}$)($d^2N/dp_{\perp}dy$)
and $<N_{coll}>$ is defined as above.

The scaling behaviour of the pions is different 
from those of the sum of protons and anti-protons.
The pions yield scaled by $N_{coll}$ in central events is
strongly suppressed compared to 
pp reactions ($R_{AA}$) and to peripheral events ($R_{cp}$).
The hadron production  in  HIJING/B\=B v2.0 is mainly from the  
fragmentation of energetic partons. Thus, 
the observed suppression in central collisions may be
a signature of the energy loss of partons 
during their propagation through the hot and dense matter (possibly QGP)
created in the collisions, i.e {\em jet quenching}.

HIJING/B\=B v2.0 predicts that the sum of protons and anti-protons 
(p+\=p)/2, scale wells with the number of 
binary collisions ($N_{coll}$).
On the other hand the discrepancy seen at $p_T<1.5\, $ GeV/c
indicates a sizeable contribution from radial flow. 
Similar trend are observed in $\Lambda$, $K_S^0$ and $K^{\pm}$
measurements by the STAR Collaboration \cite{Adams:2003am}.
It has been recently shown that the competition between 
recombination and parton fragmentation  
at moderate $p_T$ may also explain  \cite{lin02} the observed features.

%%%%%%%%%%out of text result remarks

%%In particular will we explore interplay of two distinct effects:

%%i) initial state Cronin enhancement 

%%ii) SCF effects on string tension and intrinsic transverse momentum

%%iii) jet quenching and shadowing effects.

%%%%%%%%%%%%%%%%%%%%%%%%%%%%%%%%%%%%%%%%%%%%%%%%%%%%%%%

\section{Summary and Conclusions}

The failure of the previous implementation of baryon junction loops
in HIJING/B\=B v1.10 to reproduce the observed $p_T$ enhancement
of anti-baryons and baryons motivated us to construct 
a new version,  HIJING/B\=B v2.0.

The new physics ingredients implemented in  HIJING/B\=B v2.0, 
concentrate on modifications of the hypothesized  
junction J\=J loops algorithm as well as adding a
phenomenological simulation of collective transverse flow.
%final state interactions. 
These modifications make a significant improvement
in  the full event Monte Carlo description of 
a large set of observables for p and \=p. The new version
can account now for many features of the baryon anomaly region
at moderate $p_T$,
 as well as for characteristic stopping observables at 
$\sqrt{s_{NN}} $=200 GeV in Au+Au collisions.

A simultaneously absence of suppression for baryons up to 
$p_T$=4-5 GeV/c and the enhancement of the $p/\pi$ ratios
at moderate $p_T$ which is a challenge for many theoretical 
framework is well described within HIJING/B\=B v2.0
with shadowing and jet quenching effects, for an energy loss parameter 
dE/dx =1 GeV/fm (for quark jet) and 
a constant phenomenological factor $F_{p_T}$=3.  
One of the remaining discrepancies is the energy loss per 
participant nucleon and 
baryon rapidity measurements at forward rapidity ( $y> 3$),
which will require futher analysis.
%in order to use this observable as a signature
%for partonic processes and for QGP formation.

While this new version  HIJING/B\=B v2.0  gives 
a good description of a large body of data it still cannot reproduce the 
transverse mass spectra of kaons for which the integrated yield is well
predicted, but the model has no mechanism to account for kaon radial flow. 
In string fragmentation phenomenology, the strong enhancement
of strange particle observables
require strong color field effects  (SCF) \cite{soff_jpg04,nu03_scf}.  
The full understanding of the production of 
strange particles in relativistic heavy-ion collisions
\cite{bratkovskaya_04} remains an exciting open question.

%The model presented here could be an useful tool to 
%guide experiment in their Monte Carlo simulations.

\section{Acknowledgments}

{\bf Acknowledgments:} The authors would like to thank Subal Das Gupta
for careful reading of the manuscript and to Stephen Vance.
This work was partly supported by the Natural Science and 
Engineering Research Council of Canada and the 
``{\it Fonds Nature et Technologies}'' of Quebec.  
This work was supported also by the Director,
Office of Energy Research, Office of High Energy and Nuclear Physics,
Division of Nuclear Physics, and by the Office of Basic Energy
Science, Division of Nuclear Science, of the U. S. Department 
of Energy under Contract No. DE-AC03-76SF00098 and
DE-FG-02-93ER-40764.

%%%%%%%%%%%%%%%%%%%%%FIG1 \label{fig01_st} %%%%%%%%%%%%%%

%%%%%%%%%%%%%%%%%%%%%%%%%%

\newpage 

\begin{figure} [hbt!]

\vskip 0.5cm

\centering

\epsfig{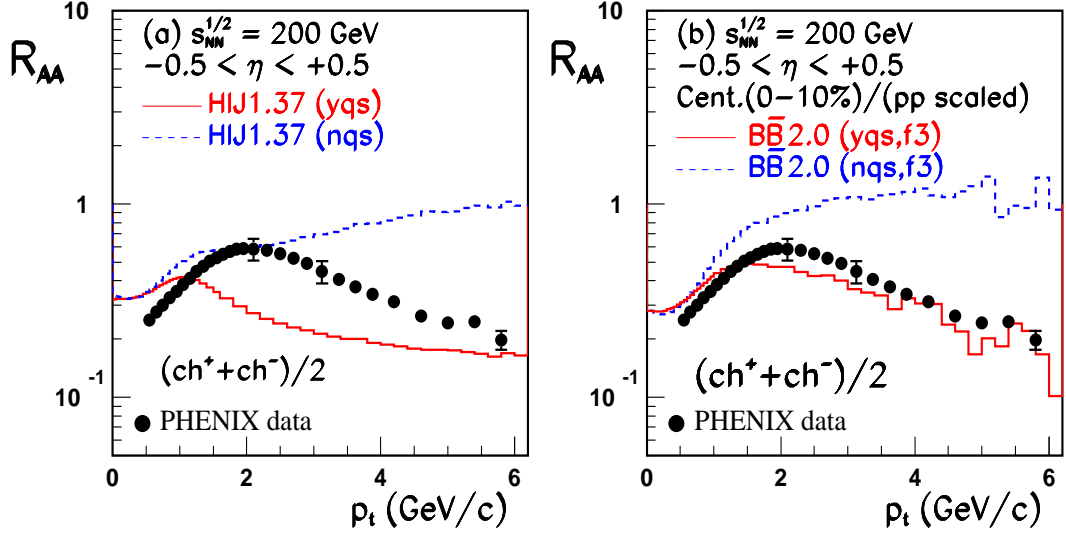} 

\vskip 0.5cm

\caption[Raa vs pt AA] {\small (Color online)
Comparison of HIJING v1.37 (left part) and HIJING/B\=B v2.0 (right part)
predictions for the nuclear modification factor ($R_{AA}$)
for central (0-10\%) Au+Au collisions at 
$\sqrt{s_{NN}}$=200 GeV. The results are 
with (solid histograms-yqs) or without (dashed histograms-nqs) 
shadowing and quenching effects included. The label f3 stands for 
model calculations assuming $F_{p_T}=3$.
The data are from PHENIX \cite{Adler:2003au}.
Only statistical error bars are shown.
The error bars at $p_T$=2 GeV and $p_T$=3 GeV,   
include systematic uncertainties.
Equivalent error bars on the other points have been 
omitted for clarity.

\label{fig:fig01_st}

}

\end{figure}

\newpage 

\begin{figure} [hbt!]

\vskip 0.5cm

\centering

\epsfig{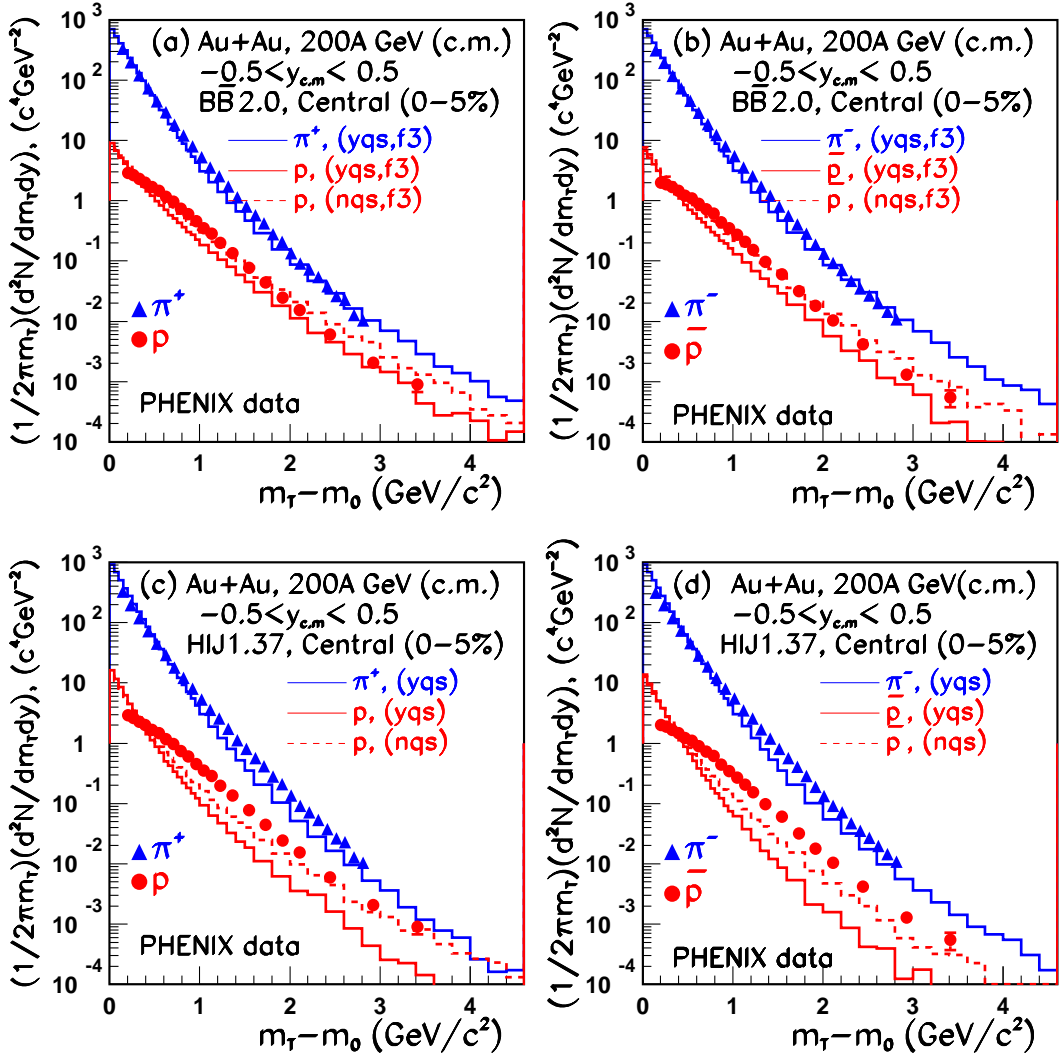} 

\vskip 0.5cm

\caption[p_t pions,protons ] {\small (Color online)
HIJING/B\=B v2.0 (upper panel) and HIJNG v1.37
(lower panel) predictions of 
transverse mass distributions for positive pions ($\pi^+$) and protons (p) 
(left part), and negative pions ($\pi^-$) and anti-protons (\=p)
(right part). The solid and dashed histograms 
have the same meaning as in Fig.~\ref{fig:fig01_st}.  
The data are from PHENIX \cite{Adler:2003cb}. The error
bars show statistical errors only.
\label{fig:fig02_st}

}

\end{figure}

\newpage

\begin{figure} [hbt!]

\vskip 0.5cm

\centering

\epsfig{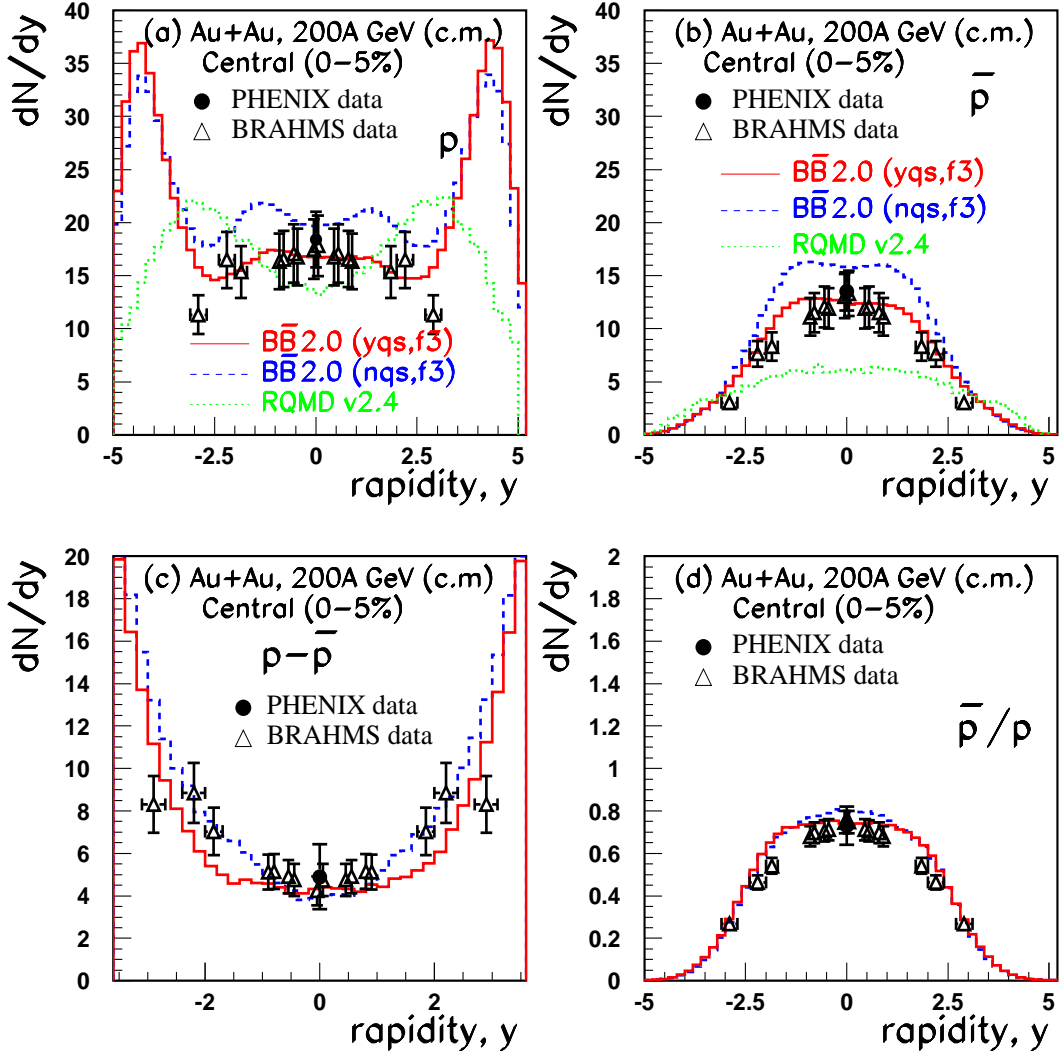} 

\vskip 0.5cm

\caption[p and anti-p] {\small (Color online)
Calculated rapidity dependence of a) protons, b) anti-protons,
c) net-protons, and d) anti-proton to proton ratio. 
The solid and dashed histograms have the same meaning as 
in Fig.~\ref{fig:fig01_st}.
The dotted histograms correspond to RQMD v2.4 model predictions. 
The data, corrected for weak decays, 
are from PHENIX \cite{Adcox:2003nr} and BRAHMS
\cite{Bearden:2003hx}. 
The errors bars shown includes both statistical and 
systematic uncertainties.
\label{fig:fig03_st}

}

\end{figure}

\newpage

\begin{figure} [hbt!]

\vskip 0.5cm

\centering

\epsfig{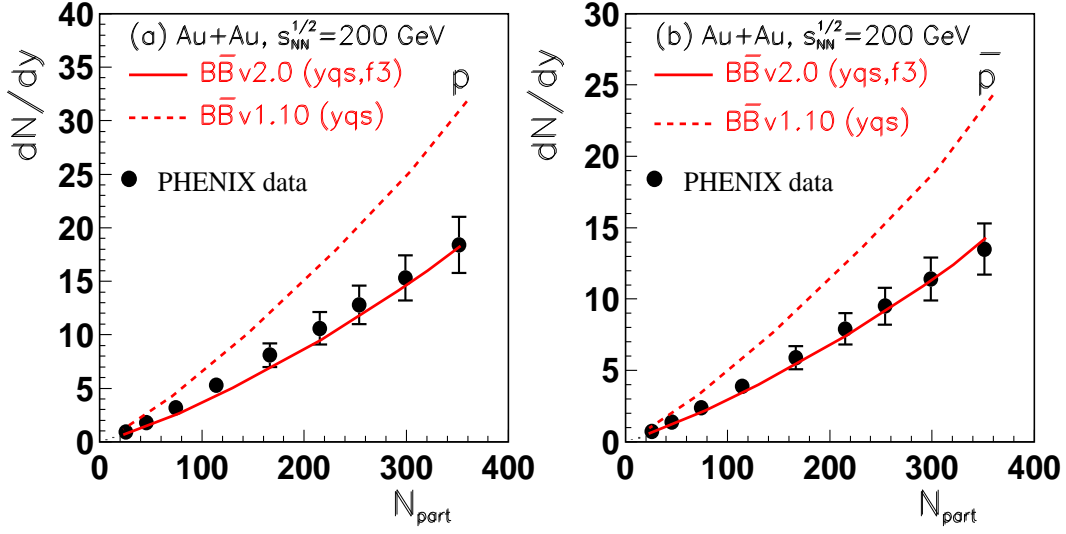} 

\vskip 0.5cm

\caption[cetrality p] {\small (Color online)
Comparison of HIJING/B\=B v1.10 (dashed lines) and 
HIJNG/B\=B v2.0 (soild lines) predictions for the centrality 
dependence of proton (part a) and anti-proton (part b)
yields at mid-rapidity.
The data are from PHENIX \cite{Adcox:2003nr}.
The errors shown includes both statistical and systematic uncertainties.
\label{fig:fig04_st}

}

\end{figure}

\newpage

\begin{figure}

\centering

\vskip 0.5cm

\epsfig{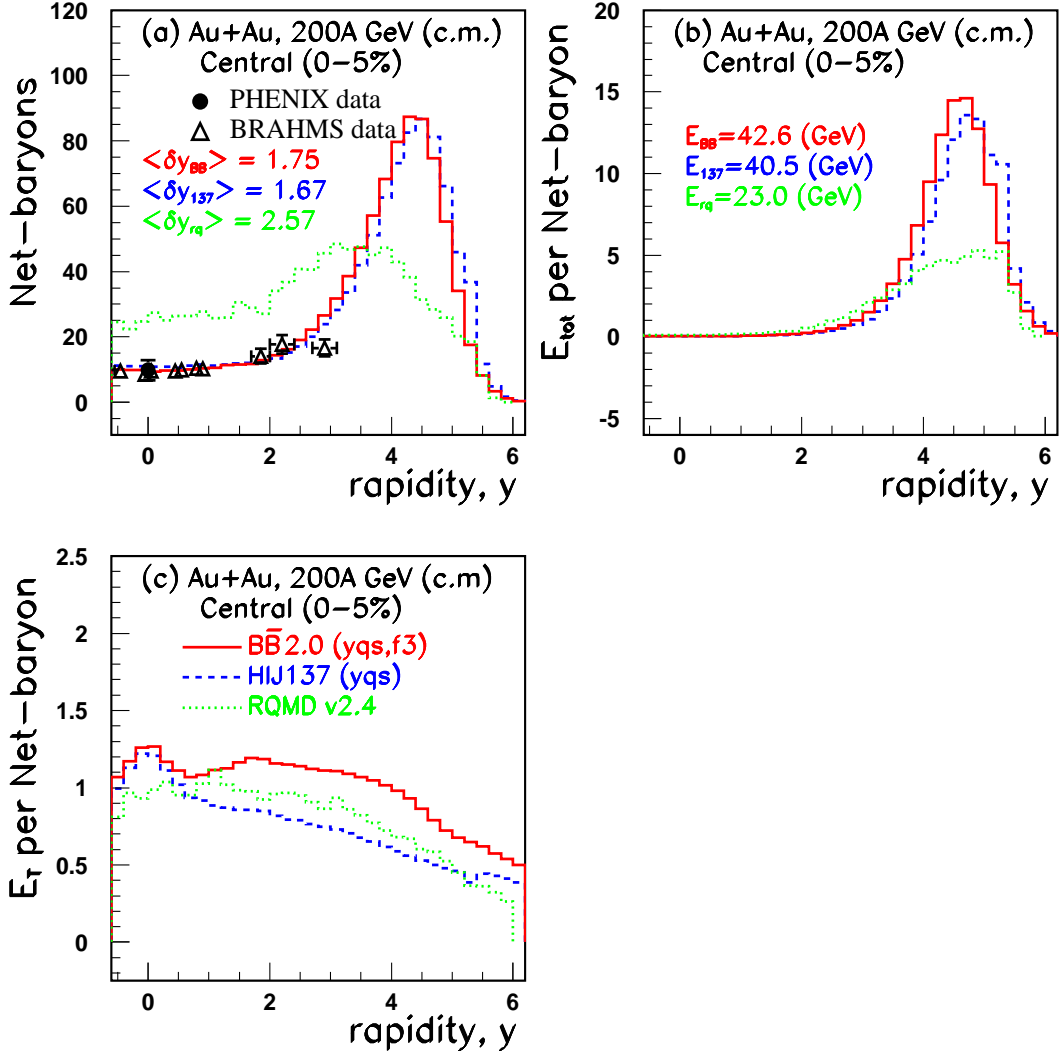}

\vskip 0.5cm

\caption[baryons distrib] {\small (Color online)
Model predictions for: a) the net-baryon
distribution and the average rapidity loss $<\delta y>$; b) 
the total energy per net-baryon after the
collisions; and c) the transverse energy per net-baryon 
for central (0-5\%) Au+Au collisions at $\sqrt{s_{NN}}$=200 GeV.
The solid and dashed histograms are the results obtained  
within HIJING/B\=B v2.0 and HIJING v1.37, respectively.
The dotted histograms are the RQMD v2.4 predictions.
The data are from BRAHMS \cite{Bearden:2003hx}.
The errors bars include both statistical and systematic uncertainties.
\label{fig:fig05_st}

}

\end{figure}

\newpage

\begin{figure}

\centering

\vskip 0.5cm

\epsfig{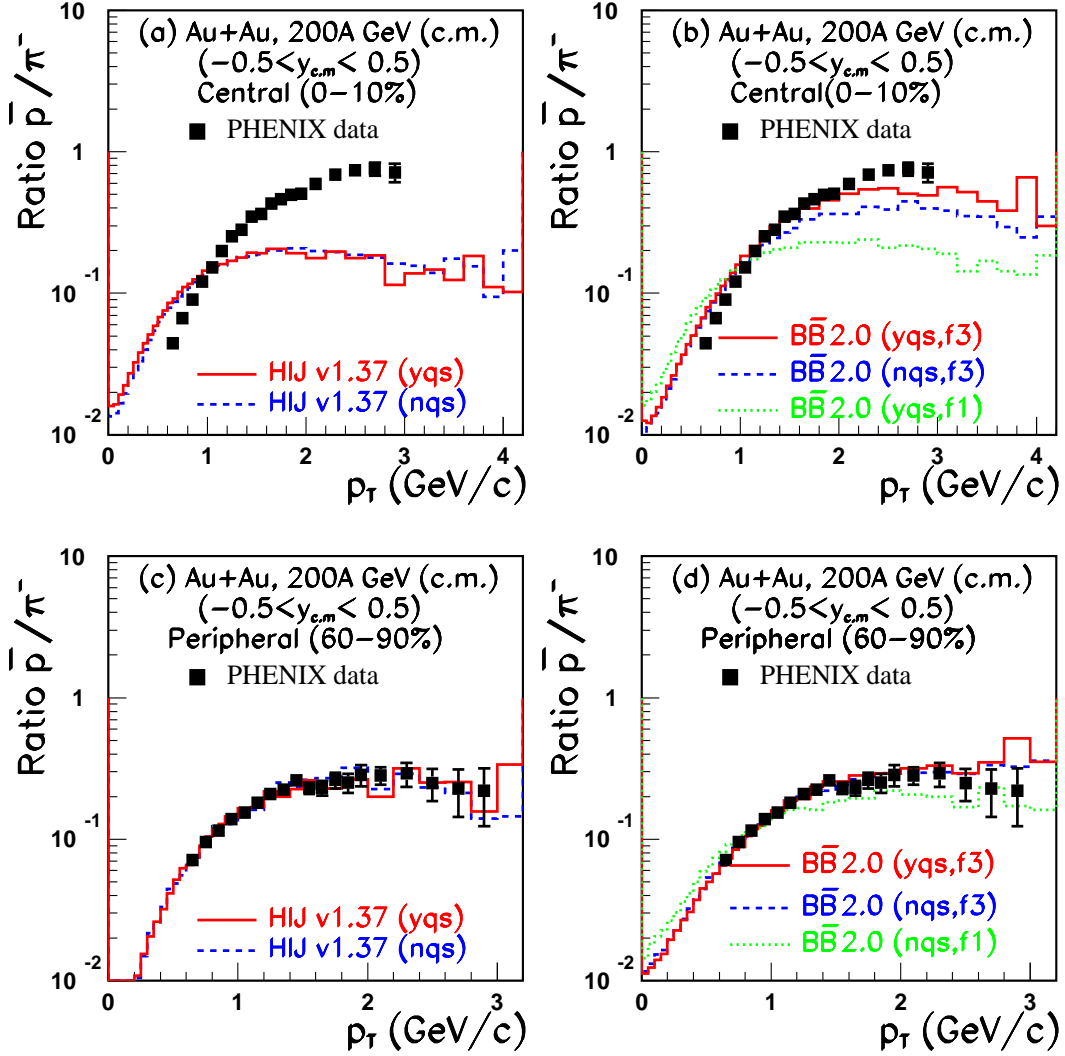}

\vskip 0.5cm

\caption[pbar/pi vs pt] {\small (Color online)
Model predictions for
$\bar{p}/\pi^-$ ratio in central (0-10 \%) Au+Au collisions at 200A GeV
(upper part) and peripheral (60-90 \%) Au+Au collisions
(lower part). 
The solid and dashed histograms have the same meaning as 
in Fig.~\ref{fig:fig01_st}. 
In figures b and d, the dotted histograms are the predictions of 
HIJING/B\=B v2.0 with $F_{p_T}$=1 (label f1). 
The data are from PHENIX \cite{Adcox:2003nr}.
The error bars include systematic uncertainties.
\label{fig:fig06_st_cp}

}

\end{figure}

\newpage

\begin{figure}

\centering

\vskip 0.5cm

\epsfig{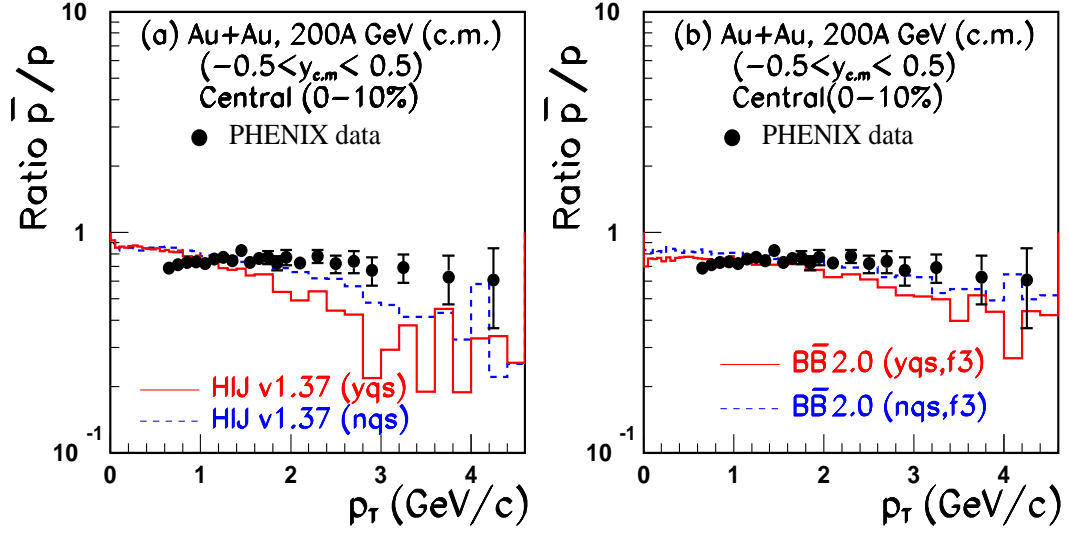}\vskip 0.5cm

\caption[pbar/pi vs pt] {\small (Color online)
Comparison of HIJING v1.37 (left) and HIJING/B\=B v2.0(right) 
model predictions for
$\bar{p}/p$ ratio versus $p_T$ for central (0-10 \%) 
Au+Au collisions at 200A GeV.
The solid and dashed histograms have the same meaning as in 
Fig.~\ref{fig:fig01_st}.
The data are from PHENIX \cite{Adcox:2003nr}.
The error bars include systematic uncertainties.
\label{fig:fig07_st}

}

\end{figure}

\newpage

\begin{figure}

\centering

\vskip 0.5cm

\epsfig{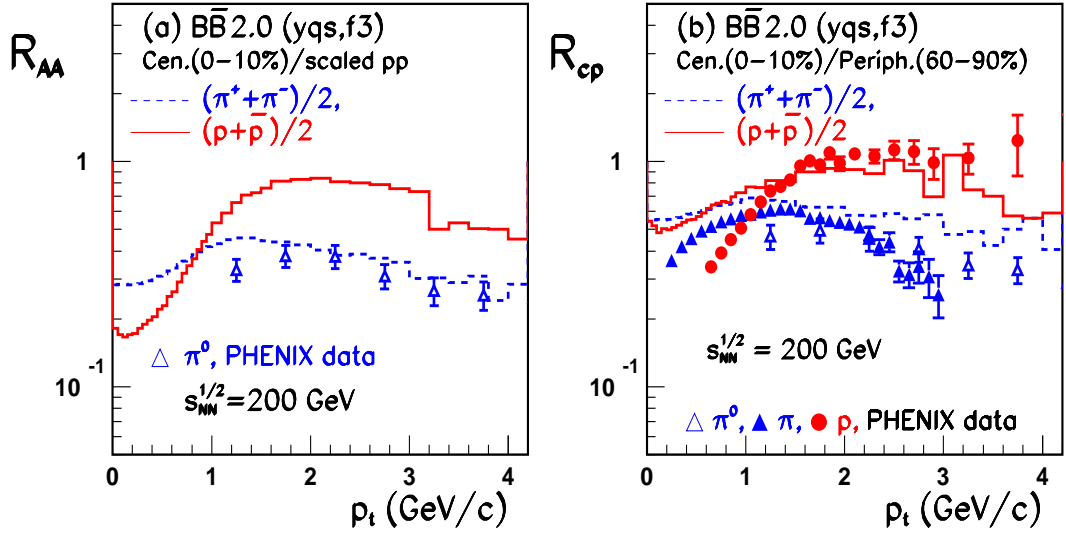}

\vskip 0.5cm

\caption[R_aa, R_cp vs pt] {\small (Color online)
HIJING/B\=B v2.0 predictions for binary-collision scaled 
%$p_T$ spectra  
nuclear modification factor $R_{AA}$ (part a)
and $R_{cp}$ (part b) for (p+\=p)/2 (solid histograms)
and charged pions (dashed histograms) in
central (0-10 \%) Au+Au collisions.
The data are from PHENIX  \cite{Adcox:2003nr}.
The error bars are statistical only.
\label{fig:fig08_st}

}

\end{figure}


\newpage





\begin{thebibliography}{199}



%%%%%%%%%%%%%%%%%%%%%%%%%%%%out of text%%%%%%%%%%%%%%%%%%%%%%%%


\bibitem{qm02} Proceedings of the ``16th International
Conference on Ultra-Relativistic Nucleus-Nucleus Collisions (QM02)'',
Nantes, France; 18-24 July, 2002; 
edited by H. Gutbrod, J. Aichelin and K. Werner; 
Nucl. Phys. {\bf A715}, 1c (2003).

\bibitem{qm04} Proceedings of ``17th International Conference on
Ultra-Relativistic Nucleus-Nucleus Collisions (QM04)''
Oakland, US; 11-17 January, 2004; J. of Phys. G (to be published).

\bibitem{gyu_qm04} M. Gyulassy, Proceedings of NATO Advanced 
Study Institute: {\it Structure and Dynamics of Elementary Matter}
(Kemer, 22 Sep.- 2 Oct.,2003), edited by W. Greiner
; nucl-th/0403032 (to be published).

\bibitem{bnl04} Proceedings of RIKEN BNL Workshop ``{\it New Discoveries
at RHIC-The Strongly Interactive QGP}'', 14-15 May 2004, Brookhaven, US;
RBRC Scientific Articles, Volume {\bf 9}, 1 (2004).

\bibitem{gyu_bnl04} M. Gyulassy and L. McLerran,
ibidem \cite{bnl04}, p23 (2004).

\bibitem{muller04} B. M$\ddot{u}$ller, 
ibidem \cite{bnl04} , p77 (2004).

%%%%%%%%%%%%%%%%%%%%%%%%%%%%%%%%%%%%%%%%%%%

%%%%%%%%\bibitem{block_exp} To introduce exp work%%%%%%

%%%%%%%%%%%%%%%%%%%%%%%%%%%%%%%%%%%%%%%%%%%

\bibitem{Adcox:2001mf} K. Adcox  {\it et al.}, [PHENIX Collaboration],
Phys. Rev. Lett. {\bf 88}, 242301 (2002);
Phys. Rev. Lett. {\bf 89}, 092302 (2002).

%%@Article{Adcox:2001mf,
%%     author    = "Adcox, K. and others",
%% collaboration = "PHENIX",
%%     title     = "Centrality dependence of pi+-, K+-, p and anti-p production
%%                  from  s(NN)**(1/2) = 130-GeV Au + Au collisions at RHIC",
%%     journal   = "Phys. Rev. Lett.",
%%     volume    = "88",
%%%     year      = "2002",
%%     pages     = "242301",
%%     eprint    = "nucl-ex/0112006",
%%     SLACcitation  = "%%CITATION = NUCL-EX 0112006;%%"
%%%%%%%%%%%%%%%%%%%%%%%%%%%%%%%%%%%%%%%%%%%%%%%%%%%%%%%%%%%%%%%%%



\bibitem{Adler:2003kg} S. S. Adler  {\it et al.}, [PHENIX Collaboration],
Phys. Rev. Lett. {\bf 91}, 172301 (2003). 
%%@Article{Adler:2003kg,
%%     author    = "Adler, S. S. and others",
%% collaboration = "PHENIX",
%%     title     = "Scaling properties of proton and anti-proton production in
%%                  s(NN)**(1/2) = 200-GeV Au + Au collisions",
%%     journal   = "Phys. Rev. Lett.",
%%     volume    = "91",
%%     year      = "2003",
%%     pages     = "172301",
%%     eprint    = "nucl-ex/0305036",
%%     SLACcitation  = "%%CITATION = NUCL-EX 0305036;%%"
%%%%%%%%%%%%%%%%%%%%%%%%%%%%%%%%%%%%%%%%%%%%%%%%%%%%%%%%%





\bibitem{Vitev:2001zn}
I.~Vitev and M.~Gyulassy,
%``Jet quenching and the anti-p >= pi- anomaly at RHIC,''
Phys.\ Rev.\ C {\bf 65}, 041902 (2002);
%%CITATION = NUCL-TH 0104066;%%
%hep-ph/0108045.
%\cite{Vitev:2001td}
%\bibitem{Vitev:2001td}
I.~Vitev, M.~Gyulassy and P.~Levai,
%``The role of jet quenching in the anti-p >= pi- anomaly at RHIC,''
%, Budapest 2001, High energy physics* hep2001/243,
hep-ph/0109198 (2001) .
%%CITATION = HEP-PH 0109198;%%




%\cite{Gyulassy:2003mc}
\bibitem{Gyulassy:2003mc}  M.~Gyulassy, I.~Vitev, ~X.~-N.~Wang
and ~B.~-W.~Zhang,
%{\it Jet Quenching and Radiative Energy Loss 
%in Dense Nuclear Matter}; 
in ``Quark Gluon Plasma 3'', pp123-191,
edited by ~R. ~C. ~Hwa and ~X.~-N. ~Wang (World Scientific, Singapore, 2003),
nucl-th/0302077, and references therein.



%\cite{Xu:2001zj}
\bibitem{Xu:2001zj} N.~Xu and M.~Kaneta,
%``Hadron freeze-out conditions in high energy nuclear collisions,''
Nucl.\ Phys.\ {\bf A698}, 306c (2002).
%%[arXiv:nucl-ex/0104021].
%%CITATION = NUCL-EX 0104021;%%


\bibitem{lin02} Z. W. Lin and C. M. Ko, Phys. Rev. Lett. {\bf 89},
202302 (2002); C. Nonaka, R. J. Fries and ~S.~A.~Bass,
Phys. Lett. B {\bf 583}, 73 (2004); 
~C.~Nonaka,~B.~Muller,~M.~Asakawa,~S.~A.~Bass, \\
and ~R.~J.~Fries,~Phys.~Rev.~C {\bf 69}, 031902 (2004);
~R.~C.~Hwa and~C.~B.~Yang,~Phys.~Rev.~C {\bf 67}, 034902 (2003);
~V.~Greco,~C.~M.~Ko and ~P.~Levai,~Phys.~Rev.~Lett. {\bf 90},
202302 (2003). 




\bibitem{svance99} S. E. Vance and M. Gyulassy, Phys. Rev. Lett. {\bf 83},
1735 (1999);\\
See http://www-cunuke.phys.columbia.edu/people
/svance/hjbb.html;\\
S. E. Vance, Ph. D. thesis, Columbia University,
1999,\\
http://www-cunuke.phys.columbia.edu/people/svance/thesis.html



%\cite{Rossi:1977cy}
\bibitem{Rossi:1977cy} G.~C.~Rossi and G.~Veneziano,
%``A Possible Description Of Baryon Dynamics In Dual And Gauge Theories,''
Nucl.\ Phys.\ {\bf B123}, 507 (1977).
%%CITATION = NUPHA,B123,507;%%
\bibitem{rossi80} G. C. Rossi and G. Veneziano,
 Phys. Rep. {\bf 63}, 153 (1980).

\bibitem{kharzeev96} D. Kharzeev, Phys. Lett. B {\bf 378}, 238 (1996)

%%%%%%%%%%%%%HIJNG BbarB v1.0%%%%%%%%%%%%%%%%

\bibitem{svance98} S. E. Vance, M. Gyulassy and ~X.~-N.~Wang, 
Phys. Lett. B {\bf 443}, 45 (1998).

%\cite{Kopeliovich:1988qm}

\bibitem{Kopeliovich:1988qm}
B.~Z.~Kopeliovich and B.~G.~Zakharov,
%``Novel Mechanisms Of Baryon Number Flow Over Large Rapidity Gap,''
Phys. Lett. B {\bf 211}, (1988); Z.\ Phys.\ C {\bf 43}, 241 (1989).
%%CITATION = ZEPYA,C43,241;%%

\bibitem{capella02} G. H. Arakelian, A. Capella, A. B. Kaidalov
and ~Yu.~M. ~Shabelski,~Eur.~Phys.~J. {\bf C26}, 81 (2002). 


\bibitem{kopelovic99} B. Kopeliovich and B. Povh, Phys. Lett. B 
{\bf 446}, 321 (1999).

\bibitem{bopp04} F. Bopp and Yu. M.~Shabelski,
hep-ph/0406158.


%%%%%%%%%%%%%%%%%%%%%% PHENIX %%%%%%%%%%%%%%%%%%%%%%%%%%%%%

%%%%%%CITAT%%%%%%%%%%

\bibitem{Adler:2003cb} S. S. Adler {\it et al.}, [PHENIX Collaboration],
Phys. Rev. C {\bf 69}, 034909 (2004).
%%@Article{Adler:2003cb,
%%     author    = "Adler, S. S. and others",
%% collaboration = "PHENIX",
%%     title     = "Identified charged particle spectra and yields in Au + Au
%%                  collisions at s(NN)**(1/2) = 200-GeV",
%%     year      = "2003",
%%     eprint    = "nucl-ex/0307022",
%%     SLACcitation  = "%%CITATION = NUCL-EX 0307022;%%"

%%%%%%%%%%%%%%%%%%%%%%%%%%%%%%%%%%%%%%%%%%%%%%%%%%%%%%%%%%%%%



%%%%%%CITAT%%%%%%%%%%%%

\bibitem{Adcox:2003nr} K. Adcox {\it et al.}, [PHENIX Collaboration],
Phys. Rev. C {\bf 69}, 024904 (2004).
%%%@Article{Adcox:2003nr,
%%     author    = "Adcox, K. and others",
%% collaboration = "PHENIX",
%%     title     = "Single identified hadron spectra from s(NN)**1/2 = 130-GeV
%%                  Au + Au collisions",
%%     journal   = "Phys. Rev.",
%%     volume    = "C69",
%%     year      = "2004",
%%     pages     = "024904",
%%     eprint    = "nucl-ex/0307010",
%%     SLACcitation  = "%%CITATION = NUCL-EX 0307010;%%"
%%%%%%%%%%%%%%%%%%%%%%%%%%%%%%%%%%%%%%%%%%%%%%%%%%%%%%%%%%%%%%%%




\bibitem{Adams:2003xp} J.Adams {\it et al.}, [STAR Collaboration],
Phys. Rev. Lett. {\bf 92}, 112301 (2004). 
%%@Article{Adams:2003xp,
%%     author    = "Adams, J. and others",{\it et al.}, [STAR Collaboration],
%%     collaboration = "STAR",
%%     title     = "Identified particle distributions in p p and Au + Au
%%                  collisions at s**(1/2) = 200-GeV",
%%     journal   = "Phys. Rev. Lett.",
%%     volume    = "92",
%%     year      = "2004",
%%     pages     = "112301",
%%     eprint    = "nucl-ex/0310004",
%%     SLACcitation  = "%%CITATION = NUCL-EX 0310004;%%"
%%%%%%%%%%%%%%%%%%%%%%%%%%%%%%%%%%%%%%%%%%%%%%%%%%%%%%%%%%%%%%%







\bibitem{Adams:2003ve}  J.Adams {\it et al.}, [STAR Collaboration],
nucl-ex/0306029 (2003), Phys. Rev. Lett. (submitted).
%%@Article{Adams:2003ve,
%%     author    = "Adams, J. and others",
%% collaboration = "STAR",
%%     title     = "Rapidity and centrality dependence of proton and anti-
%%                  proton production  from Au-197 + Au-197 collisions at
%%                  s(NN)**(1/2) = 130-GeV",
%%     year      = "2003",
%%     eprint    = "nucl-ex/0306029",
%%     SLACcitation  = "%%CITATION = NUCL-EX 0306029;%%"
%%%%%%%%%%%%%%%%%%%%%%%%%%%%%%%%%%%%%%%%%%%%%%%%%%%%%%%%%%%%%%%%%%



\bibitem{Adler:2001aq} C. Adler {\it et al.}, [STAR Collaboration],
Phys. Rev. Lett. {\bf 87}, 262302 (2001);
Phys. Rev. Lett. {\bf 86}, 4778 (2001);
Phys. Rev. Lett. {\bf 89} 092301 (2002).
%%@Article{Adler:2001aq,
%%     author    = "Adler, C. and others",
%% collaboration = "STAR",
%%     title     = "Measurement of inclusive antiprotons from Au + Au
%%                  collisions at  s(NN)**(1/2) = 130-GeV",
%%     journal   = "Phys. Rev. Lett.",
%%     volume    = "87",
%%     year      = "2001",
%%     pages     = "262302",
%%     eprint    = "nucl-ex/0110009",
%%     SLACcitation  = "%%CITATION = NUCL-EX 0110009;%%"
%%%%%%%%%%%%%%%%%%%%%%%%%%%%%%%%%%%%%%%%%%%%%%%%%%%%%%%%%%%%


%%%%%%%%%%%%%%%%%%%%%%%%%%%%%%% BRAHMS %%%%%%%%%%%%%%%%

%%%%%%%%%%%%%%%%%%%%%%%%%%%%%%% BRAHMS %%%%%%%%%%%%%%%%%

\bibitem{Ouerdane:2004yw} D. Ouerdane {\it et al.},
[BRAHMS Collaboration], nucl-ex/0403049,
ibidem \cite{qm04}.
%%@Article{Ouerdane:2004yw,
%%     author    = "Ouerdane, Djamel",
%% collaboration = "BRAHMS",
%%     title     = "Rapidity dependence of charged hadron production in central
%%                  Au + Au collisions at s(NN)**(1/2) = 200-GeV with BRAHMS",
%%     year      = "2004",
%%     eprint    = "nucl-ex/0403049",
%%     SLACcitation  = "%%CITATION = NUCL-EX 0403049;%%"

%%%%%%%%%%%%%%%%%%%%%%%%%%%%%%%%%%%%%%%%%%%%%%%%%%%%%%%





%%%%%%%%%%%%%%CITAT%%%%%%%%%%%%%%%%%%%%%%%

\bibitem{Bearden:2003hx} I. G. Bearden {\it et al.}, [BRAHMS Collaboration],
nucl-ex/0312023 (2003), Phys. Rev. Lett. (submitted).
%%@Article{Bearden:2003hx,
%%     author    = "Bearden, I. G. and others",
%% collaboration = "BRAHMS",
%%     title     = "Nuclear Stopping in Au+Au Collisions at sqrt(sNN) = 200
%%                  GeV",
%%     year      = "2003",
%%     eprint    = "nucl-ex/0312023",
%%     SLACcitation  = "%%CITATION = NUCL-EX 0312023;%%"
%%%%%%%%%%%%%%%%%%%%%%%%%%%%%%%%%%%%%%%%%%%%%%%%%



\bibitem{Lee:2003iq} J. H. Lee for [BRAHMS Collaboration],
Nucl. Phys. {\bf A715}, 482c (2003).
%%@Article{Lee:2003iq,
%%     author    = "Lee, J. H.",
%% collaboration = "BRAHMS",
%%     title     = "Rapidity dependent net-proton yields in Au + Au at
%%                  s(NN)**(1/2) = 200-GeV",
%%     journal   = "Nucl. Phys.",
%%     volume    = "A715",
%%     year      = "2003",
%%     pages     = "482-485",
%%     SLACcitation  = "%%CITATION = NUPHA,A715,482;%%"
%%%%%%%%%%%%%%%%%%%%%%%%%%%%%%%%%%%%%%%%%%%%%%%%%%%%%%%%%%%

\bibitem{Bearden:2003fw} I. G. Bearden {\it et al.}, [BRAHMS Collaboration],
Phys. Rev. Lett. {\bf 90}, 102301 (2003).
%%@Article{Bearden:2003fw,
%%     author    = "Bearden, I. G. and others",
%% collaboration = "BRAHMS",
%%     title     = "Rapidity dependence of charged antihadron to hadron ratios
%%                  in Au + Au collisions at S(NN)**(1/2) = 200-GeV",
%%     journal   = "Phys. Rev. Lett.",
%%     volume    = "90",
%%     year      = "2003",
%%     pages     = "102301",
%%     SLACcitation  = "%%CITATION = PRLTA,90,102301;%%"
%%%%%%%%%%%%%%%%%%%%%%%%%%%%%%%%%%%%%%%%%%%%%%%%%%%%%%%



\bibitem{Christiansen:2002gn} P. Christiansen {\it et al.}, 
[BRAHMS Collaboration], Nucl. Phys. {\bf A721}, 239 (2003).
%%@Article{Christiansen:2002gn,
%%     author    = "Christiansen, Peter",
%% collaboration = "BRAHMS",
%%     title     = "Rapidity dependence of net-protons at s(NN)**(1/2) = 200-
%%                  GeV",
%%     journal   = "Nucl. Phys.",
%%     volume    = "A721",
%%     year      = "2003",
%%     pages     = "239-242",
%%     eprint    = "nucl-ex/0212002",
%%     SLACcitation  = "%%CITATION = NUCL-EX 0212002;%%"
%%%%%%%%%%%%%%%%%%%%%%%%%%%%%%%%%%%%%%%%%%%%%%%%%%%%%%%%%%%%%

\bibitem{Bearden:2002ry} I. G. Bearden {\it et al.}, [BRAHMS Collaboration],
Nucl. Phys. {\bf A698}, 667c (2002).
%%@Article{Bearden:2002ry,
%%     author    = "Bearden, I. G.",
%% collaboration = "BRAHMS",
%%     title     = "Anti-proton to proton ratio in s(NN)**(1/2) = 130-GeV Au Au
%%                  collisions",
%%     journal   = "Nucl. Phys.",
%%     volume    = "A698",
%%     year      = "2002",
%%     pages     = "667-670",
%%     SLACcitation  = "%%CITATION = NUPHA,A698,667;%%"
%%%%%%%%%%%%%%%%%%%%%%%%%%%%%%%%%%%%%%%%%%%%%%%%%%%%%%%%



\bibitem{Bearden:2001kt} I. G. Bearden {\it et al.}, [BRAHMS Collaboration],
Phys. Rev. Lett. {\bf 87}, 112305 (2001).
%%     @Article{Bearden:2001kt,
%%     author    = "Bearden, I. G. and others",
%% collaboration = "BRAHMS",
%%     title     = "Rapidity dependence of antiproton to proton ratios in Au +
%%                  Au  collisions at s(NN)**(1/2) = 130-GeV",
%%     journal   = "Phys. Rev. Lett.",
%%     volume    = "87",
%%     year      = "2001",
%%     pages     = "112305",
%%     eprint    = "nucl-ex/0106011",
%%     SLACcitation  = "%%CITATION = NUCL-EX 0106011;%%"
%%%%%%%%%%%%%%%%%%%%%%%%%%%%%%%%%%%%%%%%%%%%%%%%%%%%%%%%%%%%





%%%%%%%%%%%%%%%%%%%%%%%%%%%% PHOBOS%%%%%%%%%%%%%%%

%%%%%%%%%%%%%%%%%%%%%%%%%%%% PHOBOS %%%%%%%%%%%%%%%



\bibitem{Back:2003ff} B. B. Back {\it et al.}, [PHOBOS Collaboration],
nucl-ex/0309013 (2003).
%%Article{Back:2003ff,
%%     author    = "Back, B. B. and others",
%% collaboration = "PHOBOS",
%%     title     = "Centrality dependence of charged antiparticle to particle
%%                  ratios near mid-rapidity in d + Au collisions at
%%                  s(NN)**(1/2) = 200-GeV",
%%     year      = "2003",
%%     eprint    = "nucl-ex/0309013",
%%     SLACcitation  = "%%CITATION = NUCL-EX 0309013;%%"
%%%%%%%%%%%%%%%%%%%%%%%%%%%%%%%%%%%%%%%%%%%%%%%%%%%%%%%%%%%%


\bibitem{Wosiek:2002ur} B. Wosiek {\it et al.}, [PHOBOS Collaboration],
Nucl. Phys. {\bf A715}, 510c (2003).
%%@Article{Wosiek:2002ur,
%%     author    = "Wosiek, Barbara and others",
%% collaboration = "PHOBOS",
%%     title     = "Identified particles in Au + Au collisions at s(NN)**(1/2)
%%                  = 200-GeV",
%%     journal   = "Nucl. Phys.",
%%     volume    = "A715",
%%     year      = "2003",
%%     pages     = "510-513",
%%     eprint    = "nucl-ex/0210037",
%%     SLACcitation  = "%%CITATION = NUCL-EX 0210037;%%"
%%%%%%%%%%%%%%%%%%%%%%%%%%%%%%%%%%%%%%%%%%%%%%%%%%%%%%%%%%%%%%%%



\bibitem{Back:2002ks} B. B. Back {\it et al.}, [PHOBOS Collaboration],
Phys. Rev. C {\bf 67}, 021901 (2003).
%%@Article{Back:2002ks,
%%     author    = "Back, B. B. and others",
%% collaboration = "PHOBOS",
%%     title     = "Ratios of charged antiparticles to particles near mid-
%%                  rapidity in  Au + Au collisions at s(NN)**(1/2) = 200-GeV",
%%     journal   = "Phys. Rev.",
%%     volume    = "C67",
%%     year      = "2003",
%%     pages     = "021901",
%%     eprint    = "nucl-ex/0206012",
%%     SLACcitation  = "%%CITATION = NUCL-EX 0206012;%%"
%%%%%%%%%%%%%%%%%%%%%%%%%%%%%%%%%%%%%%%%%%%%%%%%%%%%%%%%%%%



\bibitem{Back:2001qr} B. B. Back {\it et al.}, [PHOBOS Collaboration],
Phys. Rev. Lett. {\bf 87}, 102301 (2001).
%%@Article{Back:2001qr,
%%     author    = "Back, B. B. and others",
%% collaboration = "PHOBOS",
%%     title     = "Ratios of charged antiparticles to particles near mid-
%%                  rapidity in  Au + Au collisions at s(N N)**(1/2) = 130-
%%                  GeV",
%%     journal   = "Phys. Rev. Lett.",
%%     volume    = "87",
%%     year      = "2001",
%%     pages     = "102301",
%%     eprint    = "hep-ex/0104032",
%%     SLACcitation  = "%%CITATION = HEP-EX 0104032;%%"
%%%%%%%%%%%%%%%%%%%%%%%%%%%%%%%%%%%%%%%%%%%%%%%%%%%%%%%%%%%%%%%









%%%%%%%%%%%%%%%%%%%%%%%%%%%%FIN EXP BLOCK

%%%%%%%%%%%%%%%%%%%%%%%%%%% FIN EXP BLOCK

%%%%%%%%%%%%%%hijing1.37 %%%%%%%%%%%%%%%%%

\bibitem{hij92_99}
~X.~-N.~Wang and ~M.~Gyulassy, Phys. Rev. D {\bf 44},
3501 (1992); ibidem D {\bf 45}, 844 (1992); M. Gyulassy and
~X.~-N.~Wang, Comput. Phys. Commun. {\bf 83}, 307 (1994); 
~X.~-N.~Wang, Phys. Rep. {\bf 280}, 287 (1997);
~X.~-N.~Wang, Nucl. Phys. {\bf A661}, 609c (1999). 
%%%%%%%%%%%%%%%%%%%%%%%%%%%%%%%%%%%%%%%%%%%%%%%

%\cite{Sjostrand:1993yb}

\bibitem{Sjostrand:1993yb}
T.~Sjostrand,
%``High-energy physics event generation with PYTHIA 5.7 and JETSET 7.4,''
Comput.\ Phys.\ Commun.\  {\bf 82}, 74 (1994).
%%CITATION = CPHCB,82,74;%%


%%%%%%\cite{ToporPop:1995cg}%%%%%%%%%%%%%%%%%%%%
\bibitem{ToporPop:1995cg}
V. Topor Pop, ~M.~Gyulassy, ~X.~-N.~Wang,~A.~Andrighetto, 
~M.~Morando,
~F.~Pellegrini, ~R.~A.~Ricci and ~G.~Segato, 
Phys. Rev. C {\bf 52}, 1618 (1995); 
 ~M. ~Gyulassy,~V. ~Topor ~Pop and ~X.~-N.~Wang, 
~Phys.~Rev. C {\bf 54}, 1498 (1996).
%%Article{ToporPop:1995cg,
%%     author    = "Topor Pop, V. and others",
%%     title     = "Strangeness enhancement in p + A and S + A interactions at
%%                  SPS energies",
%%     journal   = "Phys. Rev.",
%%     volume    = "C52",
%%     year      = "1995",
%%     pages     = "1618-1629",
%%     eprint    = "nucl-th/9504003",
%%     SLACcitation  = "%%CITATION = NUCL-TH 9504003;%%"
%%%%%%%%%%%%%%%%%%

\bibitem{Gyulassy:1997mz}

M. Gyulassy, V. Topor Pop and S. E. Vance, Heavy Ion Phys. {\bf 5}, 299
(1997) 
%%%@Article{Gyulassy:1997mz,
%%     author    = "Gyulassy, M. and Topor Pop, V. and Vance, S. E.",
%%     title     = "Baryon number transport in high-energy nuclear collisions",
%%     journal   = "Acta Phys. Hung. New Ser. Heavy Ion Phys.",
%%     volume    = "5",
%%     year      = "1997",
%%     pages     = "299",
%%     eprint    = "nucl-th/9706048",
%%     SLACcitation  = "%%CITATION = NUCL-TH 9706048;%%"
%%%%%%%%%%%%%%%%%%%%%%%%%%%%%%%%%%%%%%





\bibitem{top03_prc68} ~V.~Topor~Pop,~M.~Gyulassy,~J.~Barrette, 
~C.~Gale, ~X.~-N.~Wang, ~N.~Xu and ~K.~Filimonov, 
~Phys.~Rev.~C {\bf 68}, 054902 (2003).

%%%%%%%%%%%%%%%%%%%%%%%%%%%%%%%%%%%%%%%%%%%%%%%%%%%%




%%%%%%%%%%%%%HYdro%%%%%%%%%%%%%%%%%%%%%%%%

\bibitem{heinz03} P. F. Kolb and U. Heinz, 
%{\it Hydrodynamic 
%description of ultrarelativistic heavy-ion collisions},   
in ``Quark Gluon Plasma 3'', pp634-714,
edited by ~R.~C.~Hwa and ~X.~-N.~Wang (World Scientific, Singapore, 2003),
nucl-th/0305084, and references therein.


\bibitem{therm03} P. Braun-Munzinger, D. Magestro,~K.~Redlich and
~J.~Stachel,~Phys.~Lett.~B {\bf 518},~41 (2001);
~F.~Becatini {\it et al.},~Phys.~Rev.~C {\bf 64},~024901 (2001);
~W.~Florkowski,~W.~Broniowski and ~M.~Michalec,
~Acta ~Phys. ~Polon ~B {\bf 33}, 761 (2002)





\bibitem{Sjostrand:2004pf} T. Sjostrand and P. Z. Skands,
JHEP {\bf 3}, 53 (2004).
%%@Article{Sjostrand:2004pf,
%%     author    = "Sjostrand, T. and Skands, P. Z.",
%%     title     = "Multiple interactions and the structure of beam remnants",
%%     journal   = "JHEP",
%%     volume    = "03",
%%     year      = "2004",
%%     pages     = "053",
%%     eprint    = "hep-ph/0402078",
%%     SLACcitation  = "%%CITATION = HEP-PH 0402078;%%"

%%%%%%%%%%%%%%%%%%%%%%%%%%%%%%%%%%%%%%









\bibitem{stop_pap_01} W. Busza and R. Ledoux, Ann. Rev. Nucl. Part.
Sci. {\bf 38}, 119 (1988).

\bibitem{stop_pap_02} H. Sorge, A. von Keitz, R. Mattiello, 
H. Stocker and W. Greiner, Phys. Lett. B {\bf 243}, 7 (1990).

\bibitem{stop_pap_03} L. Frankfurt and M. Strikman, 
Phys. Rev. Lett. {\bf 66}, 2289 (1991).



\bibitem{stop_pap_04} X. -N. Wang and M. Gyulassy, 
Phys. Rev. Lett. {\bf 68}, 1480 (1992).


\bibitem{capella03} A. Capella, Acta Phys. Polon. {\bf B34}, 3331 (2003).

%@Article{Capella:2003nh,
%     author    = "Capella, A.",
%     title     = "Mechanisms of multiparticle production in heavy ion
%                  collisions at high  energy",
%     journal   = "Acta Phys. Polon.",
%     volume    = "B34",
%     year      = "2003",
%     pages     = "3331-3362",
%     eprint    = "nucl-th/0303045",
%     SLACcitation  = "%%CITATION = NUCL-TH 0303045;%%"






%%%\bibitem{greiner74} W. Scheid, H. Muller, W. Greiner,

%%%Phys. Rev. Lett. {\bf 32}, 741 (1974).

%\cite{Wang:xy}

%\bibitem{Wang:xy}
%X.~N.~Wang and M.~Gyulassy,
%``Gluon Shadowing And Jet Quenching In A + A Collisions At S**(1/2) =
%200-Gev,''
%Phys.\ Rev.\ Lett.\  {\bf 68}, 1480 (1992).
%%CITATION = PRLTA,68,1480;%%

\bibitem{munzinger99} P. Braun-Munzinger and J. Stachel, 
Phys. Lett. B {\bf 465}, 15 (1999).

\bibitem{ropes_tbiro} T. S. Biro, H. B. Nielsen and J. Knoll,
Nucl. Phys. {\bf B245}, 449 (1984).

\bibitem{ropes_sorge} H. Sorge, M. Berenguer, H. Stocker and 
W. Greiner, Phys. Lett. B {\bf 289}, 6 (1992);
H. Sorge, Phys. Rev. C {\bf 52}, 3291 (1995).



\bibitem{qmd_1} S. Scherer, M. Hofmann, M. Bleicher, L. Neise,
H. Stocker and W. Greiner, New J. Phys. {\bf 3}, 8 (2001).

\bibitem{qmd_2}  M. Hofmann, S. Scherer, M. Bleicher, L. Neise,
H. Stocker and W. Greiner, Phys. Lett. B {\bf 478}, 161 (2000).

\bibitem{urqmd_1} S. Bass {\it et al.}, Prog. Part. Nucl. Phys. 
{\bf 41}, 255 (1998).

\bibitem{urqmd_2} M. Bleicher {\it et al.}, J. Phys. G {\bf25},
1859 (1999).  

\bibitem{soff_jpg04} S. Soff, J. Phys. G: Nucl. Part. Phys. {\bf 30},
 s139 (2004).

\bibitem{nu03_scf} S. Soff, J. Randrup, H. Stocker and N. Xu, 
Phys. Lett. B {\bf 551}, 115 (2003).

\bibitem{nu04_scf} S. Soff, S. Kesavan, J. Randrup, H. Stocker and N. Xu,
nucl-th/0404005 (2004), Phys. Rev. Lett. (submitted).

\bibitem{csernai01} V. Magas, L. Csernai and D. Strottman,
Phys. Rev. C {\bf 64}, 014901 (2001).



%%%%%%%%%%%%%%%%%%%%%%%%%%%%LATER%%%%%%%%%%%%%%%%%%%%%%%%

%%%%%%%%%%%%%%%%LATER CITTED%%%%%%%%%%%%%%%%%%%%%
\bibitem{Adler:2003au} S. S. Adler {\it et al.}, [PHENIX Collaboration],
Phys. Rev. C {\bf 69}, 034910 (2004).
%%%@Article{Adler:2003au,
%%%     author    = "Adler, S. S. and others",
%%% collaboration = "PHENIX",
%%%     title     = "High-p(T) charged hadron suppression in Au + Au collisions
%%%                  at s(NN)**(1/2) = 200-GeV",
%%%     year      = "2003",
%%%     eprint    = "nucl-ex/0308006",
%%%     SLACcitation  = "%%CITATION = NUCL-EX 0308006;%%"
%%%%%%%%%%%%%%%%%%%%%%%%%%%%%%%%%%%%%%%%%%%%%%%%%%%%%%%%%%


\bibitem{bass_st_03} S. A. Bass, B. Muller and D. K. Srivastava,
Phys. Rev. Lett. {\bf 91}, 052302 (2003).

\bibitem{wolschin_03} G. Wolschin, Phys. Lett. B {\bf 569}, 67 (2003).

\bibitem{Adams:2003am} J.Adams {\it et al.}, [STAR Collaboration],
Phys. Rev. Lett. {\bf 92}, 052302 (2004).
%@Article{Adams:2003am,
%%     author    = "Adams, J. and others",
%% collaboration = "STAR",
%%     title     = "Particle type dependence of azimuthal anisotropy and
%%                  nuclear modification of particle production in Au + Au
%%                  collisions at s(NN)**(1/2) = 200-GeV",
%%     journal   = "Phys. Rev. Lett.",
%%     volume    = "92",
%%     year      = "2004",
%%     pages     = "052302",
%%     eprint    = "nucl-ex/0306007",
%%     SLACcitation  = "%%CITATION = NUCL-EX 0306007;%%"
%%%%%%%%%%%%%%%%%%%%%%%%%%%%%%%%%%%%%%%%%%%%%%%%%%%%%%%%%%%



%%\bibitem{pdbook_02} Particle Data Group (K. Hagiwara {\it et al.}),
%%Phys. Rev. D {\bf66}, 010001 (2002).




%%%%%%%%%%%%%%%%%%%Strangenes E.Bratkovskaya%%%%%%%%%%%%%%%

\bibitem{bratkovskaya_04} E. L. Bratkovskaya, M. Bleicher, M. Reiter,
S. Soff, H. Stocker,~M.~van~Leeuwen,~S.~A.~Bass and ~W.~Cassing,
~Phys.~Rev. C {\bf 69}, 054907 (2004). 

%%%%%%%%%%%%%%%%%%%%%%%%%%%%%%%%%%%%%%%%%%%%%%%%%%%%%%

%%%%%%%%%%%%%%%%%%%%%%%%%%NOT CITTED%%%%%%%%%%%%%%%%%%%%



%\bibitem{Adams:2003qm} J.Adams {\it et al.}, [STAR Collaboration],
%nucl-ex/0309012, (2003). 
%%@Article{Adams:2003qm,
%%     author    = "Adams, J. and others",
%% collaboration = "STAR",
%%     title     = "Pion, kaon, proton and anti-proton transverse momentum
%%                  distributions from p + p and d + Au collisions at
%%                  s(NN)**1/2 = 200-GeV",
%%     year      = "2003",
%%     eprint    = "nucl-ex/0309012",
%%     SLACcitation  = "%%CITATION = NUCL-EX 0309012;%%"

%%%%%%%%%%%%%%%%%%%%%%%%%%%%%%%%%%%%%%%%%%%%%%%%%%%%%%%%%%%%%%%%%%


%%\bibitem{Adams:2003im} J.Adams {\it et al.}, [STAR Collaboration],
%%Phys. Rev. Lett. {\bf 91}, 072304 (2003).
%%@Article{Adams:2003im,
%%     author    = "Adams, J. and others",
%% collaboration = "STAR",
%%     title     = "Evidence from d + Au measurements for final-state
%%                  suppression of high  p(T) hadrons in Au + Au collisions at
%%                  RHIC",
%%     journal   = "Phys. Rev. Lett.",
%%     volume    = "91",
%%     year      = "2003",
%%     pages     = "072304",
%%     eprint    = "nucl-ex/0306024",
%%     SLACcitation  = "%%CITATION = NUCL-EX 0306024;%%

%%%%%%%%%%%%%%%%%%%%%%%%%%%%%%%%%%%%%%%%%%%%%%%%%%%%%%%%%%

%%\bibitem{Barannikova:2002qw} O. Barannikova, W. Fuqiang, [STAR Collaboration]%%,
%%Nucl. Phys. {\bf A715}, 458 (2003).
%%@Article{Barannikova:2002qw,
%%     author    = "Barannikova, Olga and Wang, Fuqiang",
%% collaboration = "STAR",
%%     title     = "Mid-rapidity pi+-, K+-, and anti-p spectra and particle
%%                  ratios from STAR",
%%     journal   = "Nucl. Phys.",
%%     volume    = "A715",
%%    year      = "2003",
%%     pages     = "458-461",
%%     eprint    = "nucl-ex/0210034",
%%     SLACcitation  = "%%CITATION = NUCL-EX 0210034;%%"
%%%%%%%%%%%%%%%%%%%%%%%%%%%%%%%%%%%%%%%%%%%%%%%%%%%%%%%%%%%%%%%%%%%

%%\bibitem{Xu:2002zx} Z. Xu, [STAR Collaboration],
%%nucl-ex/0207019, (2002).
%%@Article{Xu:2002zx,
%%     author    = "Xu, Zhang-bu",
%% collaboration = "STAR",
%%     title     = "Soft particle spectra at STAR",
%%     year      = "2002",
%%     eprint    = "nucl-ex/0207019",
%%     SLACcitation  = "%%CITATION = NUCL-EX 0207019;%
%%%%%%%%%%%%%%%%%%%%%%%%%%%%%%%%%%%%%%%%%%%%%%%%%%%%%%%%%%

%%\bibitem{Huang:2002rx} H. Z. Huang, [STAR Collaboration],
%% Nucl. Phys. {\bf A698}, 663 (2002).
%%@Article{Huang:2002rx,
%%     author    = "Huang, H. Z.",
%% collaboration = "STAR",
%%     title     = "Anti-baryon to baryon ratios in Au + Au collisions at
%%                  s(NN)**(1/2) = 130-GeV",
%%     journal   = "Nucl. Phys.",
%%     volume    = "A698",
%%     year      = "2002",
%%     pages     = "663-666",
%%     SLACcitation  = "%%CITATION = NUPHA,A698,663;%%"
%%%%%%%%%%%%%%%%%%%%%%%%%%%%%%%%%%%%%%%%%%%%%%%%%%%%%%%%%%

%%\bibitem{Adler:2003ii} S. S. Adler {\it et al.}, [PHENIX Collaboration],
%%Phys. Rev. Lett. {\bf 91}, 072303 (2003). 
%%@Article{Adler:2003ii,
%%     author    = "Adler, S. S. and others",
%% collaboration = "PHENIX",
%%     title     = "Absence of suppression in particle production at large
%%                  transverse  momentum in s(NN)**(1/2) = 200-GeV d + Au
%%                  collisions",
%%     journal   = "Phys. Rev. Lett.",
%%     volume    = "91",
%%     year      = "2003",
%%     pages     = "072303",
%%     eprint    = "nucl-ex/0306021",
%%     SLACcitation  = "%%CITATION = NUCL-EX 0306021;%%"

%%%%%%%%%%%%%%%%%%%%%%%%%%%%%%%%%%%%%%%%%%%%%%%%%%%%%%%%%%%%%%



%%\bibitem{Arsene:2003yk} I. Arsene {\it et al.}, [BRAHMS Collaboration], 
%%Phys. Rev. Lett. {\bf 91}, 072305 (2003).
%%@Article{Arsene:2003yk,
%%     author    = "Arsene, I. and others",
%% collaboration = "BRAHMS",
%%     title     = "Transverse momentum spectra in Au + Au and d + Au
%%                  collisions at s(NN)**(1/2) = 200-GeV and the pseudorapidity
%%                  dependence of high p(T) suppression",
%%     journal   = "Phys. Rev. Lett.",
%%     volume    = "91",
%%     year      = "2003",
%%     pages     = "072305",
%%     eprint    = "nucl-ex/0307003",
%%     SLACcitation  = "%%CITATION = NUCL-EX 0307003;%%"

%%%%%%%%%%%%%%%%%%%%%%%%%%%%%%%%%%%%%%%%%%%%%%%%%%%%%%%%


%%%\bibitem{Adams:2003kv} J.Adams {\it et al.}, [STAR Collaboration],
%%Phys. Rev. Lett. {\bf 91}, 172302 (2003).
%%Article{Adams:2003kv,
%%     author    = "Adams, J. and others",
%% collaboration = "STAR",
%%     title     = "Transverse momentum and collision energy dependence of high
%%                  p(T) hadron  suppression in Au + Au collisions at
%%                  ultrarelativistic energies",
%%     journal   = "Phys. Rev. Lett.",
%%     volume    = "91",
%%     year      = "2003",
%%     pages     = "172302",
%%     eprint    = "nucl-ex/0305015",
%%     SLACcitation  = "%%CITATION = NUCL-EX 0305015;%%"
%%%%%%%%%%%%%%%%%%%%%%%%%%%%%%%%%%%%%%%%%%%%%%%%%%%%%%%%%%%%%%%%%



%%\bibitem{Adler:2003qi} S. S. Adler  {\it et al.}, [PHENIX Collaboration],
%%Phys. Rev. Lett. {\bf 91}, 072301 (2003).
%%@Article{Adler:2003qi,
%%     author    = "Adler, S. S. and others",
%% collaboration = "PHENIX",
%%     title     = "Suppressed pi0 production at large transverse momentum in
%%                  central  Au + Au collisions at s(NN)**(1/2) = 200-GeV",
%%     journal   = "Phys. Rev. Lett.",
%%     volume    = "91",
%%     year      = "2003",
%%     pages     = "072301",
%%     eprint    = "nucl-ex/0304022",
%%     SLACcitation  = "%%CITATION = NUCL-EX 0304022;%%"
%%%%%%%%%%%%%%%%%%%%%%%%%%%%%%%%%%%%%%%%%%%%%%%%%%%%%%%%%%%%%%%%%%%


%%%%%\bibitem{star_netp1}  C. Adler {\it et al.}, [STAR Collaboration],
%%%Phys. Rev. Lett. {\bf
%%%87}, 262302 (2001); Phys. Rev. Lett. {\bf 86}, 4778 (2001);
%%%%Phys. Rev. Lett. {\bf 89} 092301 (2002).

%%\bibitem{phob_netp1} B. B. Back {\it et al.}, [PHOBOS Collaboration],
%%Phys. Rev. Lett. {\bf 87}, 102301 (2001).

%%%\bibitem{brahms_netp1} I. G. Bearden  {\it et al.}, [BRAHMS
%%Collaboration], Phys. Rev. Lett. {\bf 87}, 112305 (2001).

%%\bibitem{phe_netp1} K. Adcox {\it et al.}, [PHENIX Collaboration],
%%Phys. Rev. Lett. {\bf 88}, 242301 (2002); Phys. Rev. Lett. {\bf 89},
%%092302 (2002).





\end{thebibliography}
\end{document}